\documentclass[aps,pra,twocolumn,showpacs,superscriptaddress,10pt,amsmath,letterpaper]{revtex4-1}
\usepackage{amsmath}
\usepackage{amssymb}
\usepackage[english]{babel}
\usepackage{cancel}
\usepackage{color}
\usepackage{comment}
\usepackage{framed}
\usepackage{graphicx}
\usepackage[colorlinks,urlcolor=ora,citecolor=ora,linkcolor=azz]{hyperref}
\usepackage{natbib}
\usepackage[normalem]{ulem}
\usepackage{times,txfonts}
\definecolor{azz}{rgb}{0,0.5,1}
\definecolor{ora}{rgb}{0.1,0.7,0.2}
\newcommand{\ket}[1]{|{#1}\rangle}
\newcommand{\ketbis}[1]{{#1}\rangle}
\newcommand{\bra}[1]{\langle{#1}|}

\newcommand{\kets}[2]{|{#1}\rangle_{#2}\hspace*{-0.2mm}}

\newcommand{\ketbras}[3]{\ket{#1}_{#3}\hspace*{-0.mm}\bra{#2}}
\newcommand{\ketbrasbis}[3]{\ketbis{#1}_{#3}\hspace*{-0.2mm}\bra{#2}}

\newcommand{\Tr}{\mathrm{\text{Tr}}}
\renewcommand{\sec}[1]{Sec.~\ref{#1}}

\newcommand{\eq}[1]{Eq.~\eqref{#1}}

\newcommand{\ug}{\!=\!}

\newcommand{\AvSk}[2]{\mathcal{I}^{#1}\!\left(#2 \right)}
\newcommand{\AvSkCorr}[2]{\mathcal{I}^{#1}_{corr}\!\left(#2 \right)}
\makeatletter
	\renewcommand{\maketag@@@}[1]{\hbox{\m@th\normalsize\normalfont#1}}
\makeatother
\begin{document}
\title{Building versatile bipartite probes for quantum metrology}
\author{Alessandro Farace}
\affiliation{Max-Planck Institut f\"ur Quantenoptik, 85748 Garching, Germany}
\affiliation{NEST, Scuola Normale Superiore and Istituto Nanoscienze-CNR, I-56126 Pisa, Italy}
\author{Antonella De Pasquale}
\affiliation{NEST, Scuola Normale Superiore and Istituto Nanoscienze-CNR, I-56126 Pisa, Italy}
\author{Gerardo Adesso}
\affiliation{School of Mathematical Sciences, The University of Nottingham, Nottingham NG7 2RD, United Kingdom}
\author{Vittorio Giovannetti}
\affiliation{NEST, Scuola Normale Superiore and Istituto Nanoscienze-CNR, I-56126 Pisa, Italy}
\begin{abstract}
We consider bipartite systems as versatile probes for the estimation of transformations acting locally on one of the subsystems. We investigate what resources are required for the probes to offer a guaranteed level of metrological performance, when the latter is averaged over specific sets of local transformations. We quantify such a performance via the average skew information, a convex quantity which we compute in closed form for bipartite states of arbitrary dimensions, and which is shown to be strongly dependent on the degree of local purity of the probes. Our analysis contrasts and complements the recent series of studies focused on the minimum, rather than the average, performance of bipartite probes in local estimation tasks, which was instead determined by quantum correlations other than entanglement. We provide explicit prescriptions to characterize the most reliable states maximizing the average skew information, and elucidate the role of state purity, separability and correlations in the classification of optimal probes. Our results can help in the identification of useful resources for sensing, estimation and discrimination applications when complete knowledge of the interaction mechanism realizing the local transformation is unavailable, and access to pure entangled probes is technologically limited.
\end{abstract}
\maketitle
%
%
%
%
\section{Introduction}
\label{Sec:Intro}
Quantum metrology is one of the most promising branches of quantum technology and studies how to exploit the laws of quantum mechanics to improve the precision in the estimation or identification of some target parameter characterizing a quantum system of interest~\cite{Helstrom1976a,Huelga1997a,Giovannetti2006a,Escher2011a,Giovannetti2011a}. A typical estimation scenario involves three distinct phases~\cite{Giovannetti2006a}: (i) a probe system is initialized in an input state; (ii) the probe interacts with the system that encodes the parameter to be estimated; (iii) the output state of the probe is measured and compared with the input state. From the comparison, if we know the physical mechanism that governs the combined probe-target dynamics (e.g. the interaction Hamiltonian), we can deduce the value of the parameter. In general, the measurement process is affected by statistical errors, whose origin can be extrinsic (e.g.~environmental noise) or intrinsic (e.g.~Heisenberg uncertainty relations, input and output states being in general non-orthogonal and hence not distinguishable with certainty). 

To improve the precision of the estimation, several strategies can be adopted. First, we can optimize the input state of the probe so that the probe-target interaction is able to imprint the highest possible amount of information about the target parameter into the probe, i.e.~the input and output states become most distinguishable. In particular, there might be states of the probe that are left unchanged by the interaction with the measured system and are useless in this sense, so we usually want to avoid them. Second, we can repeat the measurement several times to enlarge our statistical ensemble of data and extract a sharper expectation value. This can be realized by preparing many copies of the probe and making them interact independently with the system (parallel scheme), or by making the same probe interact repeatedly with the system before extracting the information (sequential scheme). Third, we can exploit the presence of genuine quantum resources, such as quantum coherence, or quantum correlations either between the many copies of the probe or between the probe and some ancillary system that is kept as a reference, to gain advantage over purely classical strategies. In particular, it is well known that the presence of entanglement allows one to estimate a parameter encoded in a unitary dynamics (e.g.~a phase shift) with an error that scales as ${ 1/N }$ with respect to the number $N$ of collected measurements, while classical strategies can at most achieve a scaling of ${ 1/\sqrt{N} }$~\cite{Giovannetti2006a,Giovannetti2011a}.

In some specific cases of practical relevance, we may not have a complete trusted knowledge of the probe-target interaction mechanism and therefore we may find it harder to optimize the input state of the probe in order to maximize the efficiency of the estimation. For example, we could imagine a situation in which we become aware of unwanted noise sources just before we retrieve the output state, meaning that the actual transformation is different from what we expected when we prepared the probe, which is then likely to be sub-optimized. As another example, we could be asked to prepare a {\it passe-partout} probe state that must be good whenever the interaction with the measured system is described by a Hamiltonian picked at random from a given ensemble, so that we have no interest in optimizing the probe for a particular element of the ensemble. It turns out that in such and similar situations, that we may describe as instances of ``black-box'' quantum metrology, the presence of correlations gives another fundamental advantage~\cite{Girolami2013a,Girolami2014a,Adesso2014a,Farace2014b,Roga2015a}. While with a single probe system we always run the risk of preparing the probe in a state which is left unmodified by some unlucky interaction mechanism with the target system, by exploiting correlations between the probe and an ancillary system kept as a reference we can instead guarantee a minimum detection efficiency.

It is then interesting to ask the following question: Given a certain minimum efficiency that we want to achieve in a black-box quantum metrology task, what resource should we look for in our probe state? The answer has been found in several recent works~\cite{Girolami2013a,Girolami2014a,Adesso2014a,Farace2014b,Roga2015a} and in short is: {\it discord-type} correlations. These are general quantum correlations that encompass entanglement but also describe the nonclassical nature of most separable states. They have been introduced for the first time in 2001 under the name of quantum discord~\cite{Henderson2001a,Ollivier2001a} and have been the subject of extensive studies in the last decade~\cite{Modi2012a}. In particular, it has been recently shown that quantum correlations in a bipartite probe can be exploited to guarantee a minimum precision in the estimation of a local phase~\cite{Girolami2014a,Adesso2014a} or a minimum probability of detecting a remote object in a quantum illumination~\cite{Farace2014b} or quantum reading~\cite{Roga2015a} scenario. Let us stress the following fact.
While, as one could expect, pure maximally entangled states of the probe-ancilla bipartite system are still the best option for the considered tasks, entanglement is not a necessary resource in the black-box scenario. On the contrary, discord-type correlations embody the fundamental feature that provides, guarantees and quantifies a quantum over classical advantage in a vast class of metrology tasks (see also \cite{Modi2011a}). Therefore, one can also consider using ``cheaper'' separable but quantumly-correlated states~\cite{Ciccarello2012a,Ciccarello2012b}  if the required minimum precision is not too stringent, and in general if the production of pure entangled states is hindered by technological limitations.

In this paper we extend the above analysis a significant step further. As just discussed, the amount of discord-type quantum correlations in the input state of the probe is all the information that we need in order to know what the worst-case performance will be and hence guarantees a minimum estimation efficiency. However two states with the same amount of discord-type correlations are not fully equivalent resources from a general metrological point of view. Indeed, although they are characterized by the same minimum estimation efficiency, one of the two states could be better on average and thus preferable over the other, as long as the information about the system-target interaction remains partially unknown. For all practical purposes, truly versatile probes for quantum metrology should then be able to offer acceptable performances on average when employed for a broad range of tasks. Therefore, other than investigating the resources involved in determining a worst-case performance as done earlier, one should address a different key question: Given a certain {\it average} efficiency that we want to achieve in a black-box quantum metrology task, what resource should we look for in our probe state? Here, we discuss this aspect in full detail and we provide a comprehensive classification and characterization of bipartite quantum probe states in terms of their average metrological performance. Together with previous results~\cite{Girolami2013a,Girolami2014a,Adesso2014a,Farace2014b,Roga2015a}, our analysis can have a direct impact on the concrete search for optimal and versatile probe states useful for a plethora of metrological applications in realistic conditions.

To deliver a quantitative analysis, we focus here on the skew information ${ I(\rho, H)=- \Tr\left[ [\sqrt{\rho},H]^{2}\right]/2 }$, which expresses the amount of information stored in a state $\rho$ that cannot be accessed by measuring the observable $H$, due to the noncommutativity between state and observable~\cite{Wigner1963a,Luo2003a}. The skew information is one possible extension of the classical Fisher information to the quantum domain, being part of a larger family of Riemannian contractive metrics on the quantum state space \cite{Gibilisco2003a,Toth2014a}: therefore, it directly quantifies the susceptibility of a probe state $\rho$ to an infinitesimal change in a target parameter encoded in the observable $H$. If the observable acts locally on one subsystem of a bipartite state, the skew information is bounded from below by the amount of discord-type correlations in the state and its minimum value can be used in fact as a measure of discord-type correlations, defined in~\cite{Girolami2013a} as the Local Quantum Uncertainty (LQU). This quantity is closely related to other measures, such as the Interferometric Power (IP)~\cite{Girolami2014a} and the Discriminating Strength (DS)~\cite{Farace2014b}, that have a direct interpretation in terms of metrological tasks in worst-case scenarios. For example, the LQU coincides with the DS for qubit systems and gives a lower bound to the IP in general. Therefore the LQU can be interpreted as a minimum susceptibility of a bipartite state to local transformations on one subsystem, thus being relevant from a quantum estimation perspective. Moreover it is based on a simple functional, the skew information, that is typically easy to compute and serves as a good starting point for our investigation.

For arbitrary states of a generic bipartite system, we compute here the average of the skew information over specific classes of local observables acting on one subsystem. The resulting quantity, referred to as Average Skew Information (AvSk), quantifies therefore the average susceptibility of a bipartite state to local transformations. Remarkably, such an average susceptibility can be expressed through a simple analytical expression, that clearly shows what is the role played by the properties of the observables and by the properties of the state in determining the average performance.  Thanks to this, we provide an extensive characterization of the AvSk and of its features. In the specific case of a two-qubit system, where the LQU is also computable in closed form \cite{Girolami2013a}, we then carry out a parallel study of our new quantity and of the LQU that allows us to identify which states of the probe are better given different constraints. It turns out that the resources needed in the probe state to optimize the average metrological performance are quite distinct from those (discord-type correlations) needed instead to guarantee a minimum performance. We also find that our AvSk is equivalent, up to a numerical prefactor, to another quantity recently introduced by Luo and coworkers~\cite{Luo2012a} which is similarly based on the skew information but considers a different kind of averaging. This connection allows us to easily prove that the AvSk can be adapted to define a measure of correlations but not specifically of quantum (like the LQU) or classical correlations. Furthermore, our analysis complements that of Luo et al. by finding a nice closed analytic expression and a clear operational meaning for their measure. Finally, we also compute the variance of the skew information to investigate what additional knowledge can be gained from higher moments of the statistics.

The main content of the paper is structured as follows. In \sec{Sec:Average} we compute the average of the skew information over an ensemble of local observables with fixed non-degenerate spectrum. In \sec{Sec:Properties} we enumerate and prove the basic properties of the average skew information. In \sec{Sec:Spectrum} we discuss how the average skew information depends on the choice of the spectrum of the local observable. In \sec{Sec:State} we compute the average skew information for specific classes of states and we derive some general bounds. In \sec{Sec:Qubits} we make a detailed analysis of the two-qubit case, comparing the average skew information with the LQU (i.e. the minimum skew information). In \sec{Sec:Variance} we also compute the variance of the skew information and we discuss what this refined statistics can tell us about the presence of quantum correlations. In \sec{Sec:Correlations} we discuss the connection between our quantity and the one recently introduced by Luo et al. \cite{Luo2012a}, and we provide additional comments on the role of correlations. Finally, in \sec{Sec:Metrology} we provide an explicit interpretation of the main results of this paper from a metrological point of view. We present our concluding remarks in \sec{Sec:Conclusions}. Some technical derivations are deferred to Appendices.
%
%
\section{Average of the skew information over local observables with fixed non-degenerate spectrum}
\label{Sec:Average}
If $\rho$ is a density operator on a Hilbert space $\mathcal{H}_X$ and $H$ is an Hermitian operator on $\mathcal{H}_X$, the skew information of $\rho$ with respect to $H$ is defined as~\cite{Wigner1963a, Luo2003a}
	\begin{equation}
		I(\rho, H)=-\frac{1}{2} \Tr\left[ [\sqrt{\rho},H]^{2}\right],
		\label{Eq:Skew}
	\end{equation}
and expresses the amount of information stored in a state $\rho$ that cannot be accessed by measuring the observable $H$, due to the noncommutativity between state and observable. Note that in general it is always possible to find an observable $H_{\rho}$ which is diagonal in the eigenbasis of $\rho$ and therefore can grant complete knowledge of the state, i.e. ${I(\rho, H_{\rho})=0}$. However, this is no longer true if we make the additional assumption that observables act only on a part of the global system.

It has been recently shown~\cite{Girolami2013a} that when ${ \rho=\rho_{AB} }$ is a density operator of a bipartite system described by the Hilbert space ${ \mathcal{H}_{AB}=\mathcal{H}_{A}\otimes\mathcal{H}_{B} }$ and ${ H=H_{A} \otimes \mathbb{I}_{B} }$ is a local Hermitian operator acting only on $\mathcal{H}_A$, the skew information is bounded from below by the presence of general nonclassical correlations of the discord type~\cite{Henderson2001a,Ollivier2001a,Modi2012a} in the state $\rho$. Quantum discord, as proposed in the original formulation~\cite{Henderson2001a,Ollivier2001a}, measures the part of the information stored in the correlations of a bipartite system ${AB}$ that cannot be retrieved by measuring locally one of the subsystems (say $A$). This locally unaccessible correlations arise because a local measurement can perturb the state of the system by projecting it onto a particular local basis for $A$, losing some information in the process, and the existence of an unperturbing measurement is not guaranteed. In the same spirit, taking the minimum of the skew information over some ensemble of local observables of a bipartite system gives the minimum incompatibility between the state $\rho$ and the ensemble of observables, i.e. the amount of information that always remains hidden under a certain family of local measurements. In particular, if one considers the set of all local observables with a fixed non-degenerate spectrum, one obtains the LQU introduced in~\cite{Girolami2013a}
	\begin{equation}
		\mathcal{U}^{\Lambda_{A}}(\rho)=\min_{\{H(\Lambda_{A})\}} I(\rho, H(\Lambda_{A})) \;,
		\label{Eq:LQU}
	\end{equation}
which is in fact a good quantifier of discord-type correlations. In \eq{Eq:LQU} the minimum is taken over a set of local observables with fixed non-degenerate spectrum ${ \Lambda_A = \sum_{i} \lambda_{i} \ket{i}_{A} \bra{i} }$, where $\{ \ket{i}_{A} \}$ is an orthonormal basis of $A$ and the $\lambda_{i}$'s are all different. This is necessary to ensure that the identity $\mathbb{I}_{A}$ is excluded from the minimization set and the trivial case ${I(\rho, \mathbb{I}_A) = 0}$ is avoided (this must hold also if considering any subspace of $\mathcal{H}_{A}$). That is, only observables of the form ${H(\Lambda_{A}) = U_A \Lambda_A {U_A}^\dagger}$ are considered, where $U_A$ is any local unitary transformation on subsystem $A$. As shown in~\cite{Girolami2013a}, the LQU satisfies all the properties required to a well-behaved measure of discord-type quantum correlations~\cite{Brodutch2012a,Aaronson2013a}. In particular it is zero if and only if the original quantum discord is zero and hence captures the same type of correlations. Moreover, the LQU is strongly connected to other measures of quantum correlations, such as the IP~\cite{Girolami2014a} and the DS~\cite{Farace2014b}, that have a clear interpretation in a metrological context. For example, the LQU coincides with the DS if the bipartite system is made of two qubits, and in this case it measures the minimum efficiency of a given bipartite state as a probe for a quantum illumination task~\cite{Lloyd2008a} where one must decide if any transformation in a given set of isospectral local unitary operations has been performed or not on the probe.

Here, instead of taking the minimum as in \eq{Eq:LQU}, we compute the average of the skew information over the set of Hermitian operators ${U_A \Lambda_A {U_A}^\dagger}$ spanned by the unitary group on $\mathcal{H}_{A}$. In light of the above discussion, this quantity, which will be named simply {\it Average Skew Information (AvSk)}, can be interpreted as the average susceptibility of a bipartite probe to local transformations and local parameters. The AvSk can be written as an integral with respect to the Haar measure of the unitary group $d \mu_H (U_A)$
	\begin{eqnarray}	\label{Eq:Average}
		\AvSk{\Lambda_{A}}{\rho} &=& \int d \mu_H (U_A) I(\rho, U_A \Lambda_A {U_A}^\dagger) \nonumber \\
 &=& -\frac{1}{2} \int d \mu_H (U_A) \Tr\left[ [\sqrt{\rho},U_A \Lambda_A {U_A}^\dagger]^2\right].
	\end{eqnarray}
In choosing our notation, we made explicit the fact that the AvSk depends only on the state and on the specific choice of the spectrum. To compute the integral in \eq{Eq:Average} we start by rewriting \eq{Eq:Skew} for the case of a bipartite state $\rho=\rho_{AB}$ and a local observable ${H=H_{A} \otimes \mathbb{I}_{B}}$ as
	\begin{eqnarray} 		\label{Eq:Skew-swap}
		I(\rho, H_A)&=& \Tr[(\sqrt{\rho} H_A) (H_A \sqrt{\rho})-(\sqrt{\rho} H_A) (\sqrt{\rho} H_A)] \nonumber \\
				&=& \Tr[(\sqrt{\rho_{AB}} H_A \otimes H_{A'} \sqrt{\rho_{A'B'}} \nonumber \\ && \quad - \sqrt{\rho_{AB}} H_A \otimes \sqrt{\rho_{A'B'}} H_{A'})S_{AB|A'B'}],
	\end{eqnarray}
where following the procedure of Ref.~\cite{DePasquale2012a} we introduced a copy ${\mathcal{H}_{A'B'}=\mathcal{H}_{A'}\otimes\mathcal{H}_{B'}}$ of the original Hilbert space ${\mathcal{H}_{AB}=\mathcal{H}_{A}\otimes\mathcal{H}_{B}}$ and the swap operator $S_{AB|A'B'}$ acting on ${\mathcal{H}_{AB} \otimes \mathcal{H}_{A'B'}}$~\cite{Brun2001a}. Using \eq{Eq:Skew-swap} and the properties of the swap operator (see Appendix~\ref{App:Swap}) we can now rewrite \eq{Eq:Average} as
	\begin{eqnarray} \nonumber
		\AvSk{\Lambda_{A}}{\rho} &=&\Tr \Big[ \big(\rho_{AB} \otimes \mathbb{I}_{A'B'} - \sqrt{\rho_{AB}}\otimes \sqrt{\rho_{A'B'}} \big) \\&&\quad \times \; {\cal T}^{(2)} (\Lambda_{A}\otimes\Lambda_{A'}) \; S_{AB|A'B'} \Big] ,
		\label{Eq:Average-swap}
	\end{eqnarray}
where ${ {\cal T}^{(2)} (\Lambda_{A}\otimes\Lambda_{A'}) }$ is the so-called twirling channel~\cite{Horodecki1999a,Lee2003a,Vollbrecht2001a} applied to the operator ${\Lambda_{A}\otimes\Lambda_{A'}}$ (see Appendix~\ref{App:Twirling})
	\begin{eqnarray}
		{\cal T}^{(2)} (\Lambda_{A}\otimes\Lambda_{A'}) &=& \!\! \int \! d \mu_H (U_A) (U_A \otimes {U_{A'}})  (\Lambda_{A}\otimes\Lambda_{A'}) ({U^\dagger_{A}} \otimes {U^\dagger_{A'}}) \nonumber \\
											&=& \frac{N_A  \Tr[\Lambda_A ]^2  - \Tr[\Lambda_A^2 ] }{N_A(N_A^2-1)} \mathbb{I}_{A A'} \nonumber \\ &&+ \frac{N_A  \Tr[\Lambda_A^2 ]  - \Tr[\Lambda_A  ] ^2}{N_A(N_A^2-1)} S_{A|A'} .
		\label{Eq:Twirling}
	\end{eqnarray}	
In writing \eq{Eq:Twirling} we introduced the dimension $N_{A}$ of the Hilbert space $\mathcal{H}_{A}$. Plugging the last two lines of \eq{Eq:Twirling} into \eq{Eq:Average-swap}, using again the properties of the swap operator, and evaluating the trace, we finally get a remarkably compact formula for the AvSk of an arbitrary bipartite state $\rho$,
	\begin{equation}
		\AvSk{\Lambda_{A}}{\rho} =\frac{N_A  \Tr[\Lambda_A^2 ]  - \Tr[\Lambda_A  ] ^2}{N_A(N_A^2-1)} \Bigg[ N_A - \Tr_B\Big[(\Tr_{A}[\sqrt{\rho}])^2\Big]\Bigg].
		\label{Eq:Average-Final}
	\end{equation}
We stress that the analytic expression~\eq{Eq:Average-Final} holds for any dimension of the Hilbert spaces $\mathcal{H}_{A}$ and $\mathcal{H}_{B}$.
%
%
\section{Properties of the average skew information}
\label{Sec:Properties}
We discuss now some properties of the AvSk ${ \AvSk{\Lambda_{A}}{\rho} }$.\\

\textbf{Property 1a --} For any fixed spectrum, the AvSk is non-negative. This is trivially true as the skew information is non-negative and this is not changed by taking the average.\\

\textbf{Property 1b --} For any fixed non-degenerate spectrum, the AvSk is zero if and only if the state is of the form \begin{equation}\label{Eq:FreeStates}
\rho_{AB} = \frac{\mathbb{I}_{A}}{N_A} \otimes \rho_{B}.
 \end{equation}
  The proof of this is rather long and is postponed to Sec.~\ref{Sec:State}.\\

\textbf{Property 2 --} The AvSk is invariant under local unitary operations $W_A, V_B$. Indeed, consider the transformation ${\rho \rightarrow (W_A \otimes V_B) \rho (W_A^\dagger \otimes V_B^\dagger)}$ which also maps ${\sqrt{\rho}}$ into ${(W_A \otimes V_B) \sqrt{\rho} (W_A^\dagger \otimes V_B^\dagger)}$. Then, by exploiting the cyclic property of the trace in \eq{Eq:Average-Final}, it is easy to see that
	\begin{align}
		\AvSk{\Lambda_{A}}{\left(W_A \otimes V_B \right) \rho \left(W_A^\dagger \otimes V_B^\dagger \right)} =  \AvSk{\Lambda_{A}}{ \rho } .	
		\label{Eq:Invariance-U}
	\end{align}\\

\textbf{Property 3 --} The AvSk is non-increasing over all completely positive trace-preserving (CPTP) maps acting locally on $B$. To show this, let us first decompose an arbitrary local CPTP map ${\Phi_{B}(\rho_{AB})}$ as a unitary interaction with an external environment followed by a partial trace over the degrees of freedom of the environment~\cite{Nielsen2000a}
	\begin{equation}
		\Phi_{B}(\rho_{AB}) = \Tr_{E} \left[ U_{BE} (\rho_{AB} \otimes \rho_{E}) U^{\dagger}_{BE}\right],
		\label{Eq:CPT}
	\end{equation}
where we can further assume that the unitary operation involves only the subsystem $B$ and the environment, without affecting $A$. Mimicking the demonstration of Property 2, we can show that the skew information satisfies the following property
	\begin{equation}
		I(U_{BE} (\rho_{AB} \otimes \rho_{E}) U^{\dagger}_{BE}, H_{A}) = I(\rho_{AB} \otimes \rho_{E}, H_{A}),
		\label{Eq:CPT-step1}
	\end{equation}
and it is also easy to see that
	\begin{equation}
		I(\rho_{AB} \otimes \rho_{E}, H_{A}) = I(\rho_{AB}, H_{A}).
		\label{Eq:CPT-step2}
	\end{equation}
Finally, it was proven in~\cite{Lieb1973a,Wehrl1978a} that
	\begin{eqnarray}
		 I(\Phi_{B} (\rho_{AB}), H_{A}) &=& I\left( \Tr_{E} \left[ U_{BE} (\rho_{AB} \otimes \rho_{E}) U^{\dagger}_{BE}\right], H_{A}\right) \nonumber \\ &\leq& I(U_{BE} (\rho_{AB} \otimes \rho_{E}) U^{\dagger}_{BE}, H_{A}) \nonumber \\ &=&  I(\rho_{AB}, H_{A}).
		\label{Eq:CPT-stepe}
	\end{eqnarray}
Since Property 3 is true for the skew information itself, it remains true also when taking the average.\\

\textbf{Property 4 --} For pure states, the AvSk is an entanglement monotone. Indeed, given a pure state ${ \ket{\psi}_{AB}}$, we have that ${ \sqrt{ \ket{\psi}_{AB}\bra{\psi}} =\ket{\psi}_{AB}\bra{\psi} }$. Plugging this into \eq{Eq:Average-Final} leaves us with
	\begin{equation}
		\AvSk{\Lambda_{A}}{\ket{\psi}_{AB}} =\frac{N_A  \Tr[\Lambda_A^2 ]  - \Tr[\Lambda_A  ] ^2}{N_A(N_A^2-1)} \Bigg[ N_A - \Tr_B\Big[ \rho_{B}^{2} \Big]\Bigg],
		\label{Eq:Average-Pure}
	\end{equation}
where ${\rho_{B} = \Tr_{A} [ \ket{\psi}_{AB}\bra{\psi} ]}$. For pure states, a convenient measure of entanglement is provided by the generalized concurrence ${C(\left| \psi \right>_{AB})}$~\cite{Rungta2001a}, which depends only on the purity of the marginal density operators ${\Tr_{B}[\rho_{B}^{2}] = \Tr_{A}[\rho_{A}^{2}]}$ as ${ C( \ket{\psi}_{AB}) = \sqrt{2 - 2\Tr_{B}[\rho_{B}^{2}]} }$. We can then rewrite
	\begin{equation}
		\AvSk{\Lambda_{A}}{\ket{\psi}_{AB}} =\frac{N_A  \Tr[\Lambda_A^2 ]  - \Tr[\Lambda_A  ] ^2}{N_A(N_A^2-1)} \Bigg[ N_A - 1 + \frac{C(\ket{\psi}_{AB})^{2}}{2} \Bigg],
		\label{Eq:Average-Conc}
	\end{equation}
which clearly makes the AvSk an entanglement monotone. Note that, however, it cannot be considered strictly speaking as a measure of entanglement since it does not vanish on all separable (product) states; still, one can obtain a fully fledged entanglement measure on pure states by rescaling the AvSk subtracting the dimension-dependent constant in Eq.~(\ref{Eq:Average-Conc}). \\

\textbf{Property 5 --} The AvSk is convex with respect to the state. The result follows simply from the convexity of the skew information~\cite{Wigner1963a}, that is preserved by taking the average,
	\begin{equation}
		\AvSk{\Lambda_{A}}{p \rho_{1} + (1-p) \rho_{2} } \leq p \; \AvSk{\Lambda_{A}}{\rho_{1}} + (1-p) \; \AvSk{\Lambda_{A}}{\rho_{2}},
		\label{Eq:Average-Convex}
	\end{equation}
for any two states $\rho_1, \rho_2$ and a probability $0 \leq p \leq 1$. This is noteworthy, since convexity is lost instead when taking the minimum rather than the average  \cite{Girolami2013a}.\\

From Property 1, we immediately see that, at variance with the LQU, the AvSk is not a proper measure of discord-type quantum correlations~\cite{Brodutch2012a}, because in general it is different from zero when evaluated on the set of classical-quantum or classical-classical states~\cite{Modi2012a}. In fact, the AvSk can be non-zero even for completely uncorrelated states, e.g.~for any state of the form $\ket{\psi}_{A} \otimes \ket{\psi}_{B}$, meaning that it is neither a measure of classical nor total correlations. Nevertheless, we will see in Sec.~\ref{Sec:Luo} how to construct a proper measure of total correlations based on a modification of the AvSk, recovering and complementing the analysis done in Ref.~\cite{Luo2012a}. We remark however that our focus here is not to define yet another abstract measure of correlations. Instead, we are going to use the AvSk operationally as a guidance to identify optimal probe states for (black-box) quantum metrology, adopting their average performance as our figure of merit.

In the following Sections we are going to compute the AvSk for some specific classes of states and derive some general bounds, that can be straightforwardly established thanks to convexity. Finally, if we study simultaneously the AvSk and the LQU, we can point out what states are better used in quantum sensing and metrology tasks such as state discrimination and parameter estimation, depending on the rules of the game.\\
%
%
\section{Dependence of the average skew information on the spectrum}
\label{Sec:Spectrum}
The expression~\eqref{Eq:Average-Final} for the AvSk that we found at the end of \sec{Sec:Average}, explicitly factors the dependence on the spectrum of the observable and the dependence on the state. In this Section, we investigate how different choices of the spectrum relate to one another.
%
%
\subsection{Invariance under translation of the spectrum}
\label{Sec:Spectrum-Shift}
First we show that if two spectra $\Lambda_{A}$ and $\Lambda'_{A}$ are connected by a rigid shift, the two induced AvSks are equal. The rigid shift condition is expressed as ${ \Lambda'_{A}=\Lambda_{A} + \eta \mathbb{I}_{A} }$, where $\eta$ is any real number. We then have
	\begin{align}
		\Tr[\Lambda'_{A}]&=\Tr[\Lambda_{A}] + \eta N_{A}, \\ \Tr[\Lambda'_{A}{}^{2}]&=\Tr[\Lambda_{A}^{2}] + 2\eta \Tr[\Lambda_{A}] + \eta^{2} N_{A}.
		\label{Eq:Shift-Trace}
	\end{align}
Plugging the above expressions into \eq{Eq:Average-Final} and considering only the part containing the spectrum, we easily see that
	\begin{align}
		\frac{N_A  \Tr[\Lambda'_A{}^2 ]  - \Tr[\Lambda'_A  ] ^2}{N_A(N_A^2-1)} & = \frac{N_A  \Tr[\Lambda_A{}^2 ]  - \Tr[\Lambda_A  ] ^2}{N_A(N_A^2-1)}.
		\label{Eq:Shift-Invariance}
	\end{align}
This implies that ${ \AvSk{\Lambda_{A}}{\rho} = \AvSk{\Lambda_{A}+\eta \mathbb{I}_{A}}{\rho} }$, $\forall \eta$. Therefore, this allows us to simplify \eq{Eq:Average-Final} by considering only spectra with trace equal to zero.
	\begin{equation}
		\Tr[\Lambda_{A}]=0 \quad \Rightarrow \quad \AvSk{\Lambda_{A}}{\rho} =\frac{ \Tr[\Lambda_A^2 ] }{N_A^2-1} \Bigg[ N_A - \Tr_B\Big[(\Tr_{A}[\sqrt{\rho}])^2\Big]\Bigg].
		\label{Eq:Average-ZeroTrace}
	\end{equation}
%
%
\subsection{Scaling under scalar multiplication of the spectrum}
\label{Sec:Spectrum-Scalar}
Next, we consider what happens if we take a spectrum $\Lambda_{A}$ and transform it to $\eta \Lambda_{A}$ by scalar multiplication. Thanks to \eq{Eq:Average-ZeroTrace}, it is immediate to see that ${ \AvSk{\; \eta \Lambda_{A}}{\rho} = \eta^{2} \; \AvSk{\Lambda_{A}}{\rho} }$ for any value of $\eta$.
%
%
\subsection{Optimal spectrum}
\label{Sec:Spectrum-Scaling}
We can now ask which spectrum yields the highest prefactor to the AvSk. From the previous results, it is obvious that multiplication of a spectrum by a big real number can make the prefactor as big as desired. However, we want here to highlight the role played by the distribution of the eigenvalues, rather than their magnitude. We can make a fair comparison by exploiting the translation invariance and the scaling introduced above, and considering only positive spectra with unit trace (i.e. we map each spectrum to a density matrix). We then see from Eq.~(\ref{Eq:Average-Final}) that all the information about the spectrum is in the prefactor $\frac{N_A  \Tr[\Lambda_A^2 ]  - 1}{N_A(N_A^2-1)}$, which for a fixed dimension $N_A$ depends only on the spectrum purity. Therefore it is immediate to see that the best spectra are those that have ${ N_{A} - 1 }$ degenerate eigenvalues, i.e. those spectra that can be mapped into pure-state density matrices by means of rigid shifts and scalar multiplications. For example one such spectrum, taken traceless to satisfy the condition discussed in Sec.~\ref{Sec:Spectrum-Shift}, is given by ${ \Lambda_{A}=\{ (N_{A}-1)/N_{A}, -1/N_{A}, \dots, -1/N_{A} \} }$.

This means that if we want to encode some information on a state but we cannot choose the encoding basis, an almost fully degenerate spectrum allows to encode, on average, the maximum amount of information. We stress that this situation is almost opposite to what happens for the LQU~\cite{Girolami2013a} and for similar measures of quantum correlations such as the IP~\cite{Girolami2014a} and the DS~\cite{Farace2014b} that consider the worst-case performance, where it is instead believed that the optimal spectrum is harmonic~\cite{Farace2014b,Rigovacca2014a}, i.e. fully non-degenerate and with equally spaced eigenvalues. Furthermore, we see that the AvSk is non-trivial as soon as the spectrum has some different eigenvalues, i.e. as soon as $\Lambda_{A} \neq \mathbb{I}_{A}$. We don't need to impose here the stricter condition of full non-degeneracy required, for example, by the LQU.
%
%
\section{Dependence of the average skew information on the state}
\label{Sec:State}
In this Section, we study the AvSk for specific classes of states or, conversely, we look for the states that yield the maximum and the minimum AvSk given specific constraints. All the results provided here hold for any dimension of $\mathcal{H}_{A} \otimes \mathcal{H}_{B}$. Without loss of generality, we consider traceless spectra (see Sec.~\ref{Sec:Spectrum-Shift}).
%
%
\subsection{Average skew information for pure states}
\label{Sec:Average-Pure}
We start by considering pure bipartite states. As we have seen in Sec.~\ref{Sec:Properties}, the AvSk takes a simple form on the set of pure states ${\kets{\psi}{AB}}$,
	\begin{equation}
		\AvSk{\Lambda_{A}}{\ket{\psi}_{AB}}  =\frac{\Tr[\Lambda_A^2 ] }{N_A^2-1} \Bigg[ N_A - \Tr_B\Big[ \rho_{B}^{2} \Big]\Bigg],
	\end{equation}
where ${\rho_{B} = \Tr_{A} [\ket{\psi}_{AB}\bra{\psi}]}$ is the reduced state of subsystem B and $\Tr_B [ \rho_{B}^{2} ]$ is its purity, which can take values between $1/\min\{N_{A},N_{B}\}$ and $1$. Therefore we can find the following bounds for the AvSk of pure states:
	\begin{equation}
		\frac{  \Tr[\Lambda_A^2 ]  }{N_A^2-1} \Big[ N_A - 1 \Big] \leq \AvSk{\Lambda_{A}}{\ket{\psi}_{AB}} \leq \frac{  \Tr[\Lambda_A^2 ]  }{N_A^2-1} \Bigg[ N_A - \frac{1}{\min\{N_A,N_B\}} \Bigg],
	\end{equation}
\normalsize
where the upper bound is saturated by pure maximally entangled states and the lower bound is saturated by pure product states.
%
%
\subsection{Average skew information for separable states}
\label{Sec:Average-Product}
Another interesting class of states is given by separable states. Here we have no entanglement and we can investigate if the presence of discord-type quantum correlations has a specific impact on the AvSk, as it has for the LQU \cite{Girolami2013a}. We start by considering a general separable state
	\begin{equation}
		\rho_{\rm sep} = \sum_{i} p_{i} \rho_{A}^{(i)} \otimes \rho_{B}^{(i)},
		\label{Eq:Separable-Def}
	\end{equation}
where $\rho_{A}^{(i)}$ and $\rho_{B}^{(i)}$ are arbitrary density matrices of $A$ and $B$, $p_{i} > 0$ and $\sum_{i} p_{i} = 1$.	
From the convexity of the AvSk (see \sec{Sec:Properties}), we have
	\begin{align}
		0 \leq \AvSk{\Lambda_{A}}{\rho_{\rm sep}}  & \leq \sum_{i} p_{i} \; \AvSk{\Lambda_{A}}{\rho_{A}^{(i)} \otimes \rho_{B}^{(i)}} \leq \max_{\{\rho_{A} \otimes \rho_{B}\}} \AvSk{\Lambda_{A}}{\rho_{A} \otimes \rho_{B}},
		\label{Eq:Average-Separable-Convex}
	\end{align}
where in the last term we take the maximum over all product states ${ \rho=\rho_{A} \otimes \rho_{B} }$. By direct substitution in \eq{Eq:Average-ZeroTrace}, we have
		\begin{align} \nonumber
		0 &\leq \AvSk{\Lambda_{A}}{\rho_{A} \otimes \rho_{B}} = \frac{  \Tr[\Lambda_A^2 ]  }{N_A^2-1} \Big[ N_A - \Tr[\sqrt{\rho_{A}}]^{2} \Big] \\  & \leq \frac{  \Tr[\Lambda_A^2 ]  }{N_A^2-1} \Big[ N_A - 1 \Big],
		\label{Eq:Average-Product}
	\end{align}
and finally
	\begin{align}
		0 \leq \AvSk{\Lambda_{A}}{\rho_{\rm sep}} \leq \frac{  \Tr[\Lambda_A^2 ]  }{N_A^2-1} \Big[ N_A - 1\Big].
		\label{Eq:Average-Separable}
	\end{align}
The lower bound is saturated, for example, by product states of the form ${ \rho = \mathbb{I}_{A}/N_{A} \otimes \rho_{B} }$ (as announced in Sec.~\ref{Sec:Properties}, and as we are going to show, these are the only states with zero AvSk) while the upper bound is saturated, for example, by product states where the local density matrix on $A$ is pure, i.e. ${ \rho = \ket{\psi}_{A}\bra{\psi} \otimes \rho_{B} }$. 

A few remarks are in order here. First of all, we notice that all separable states yield a lower AvSk than any pure entangled state. We can then use the AvSk as a witness of entanglement and say that
	\begin{align}
		\AvSk{\Lambda_{A}}{\rho} > \frac{  \Tr[\Lambda_A^2 ]  }{N_A^2-1} \Big[ N_A - 1\Big] \quad \Rightarrow \quad \rho \text{ is entangled}.
		\label{Eq:Average-Witness}
	\end{align}	
Furthermore, since the maximum AvSk among separable states is reached by a completely uncorrelated state, we can claim that the presence of quantum correlations other than entanglement has no specific effect on the average susceptibility of a bipartite state to local transformations. Of great importance is instead the local purity of the probing subsystem $A$: as soon as $\rho_A$ is not maximally mixed, an average metrological performance is guaranteed even in absence of a correlated reference subsystem $B$.

We recall, however, that discord-type correlations as measured by the LQU determine instead the minimum susceptibility of a bipartite state to local transformations. A comparative analysis of the AvSk and of the LQU can serve then to identify states that simultaneously yield satisfactory levels of complementary figure of merits and emerge as suitable probes for sensing applications. We will come back to this point in Sec.~\ref{Sec:Qubits}, where we investigate the specific case of two qubits.
%
%
\subsubsection*{Average skew information for classical-quantum states}
\label{Sec:Average-CQ}
We compute here the AvSk for a specific class of separable states, i.e classically correlated states that have zero LQU (or equivalently zero quantum discord). Since we are considering local measurements on subsystem $A$, the set of classically correlated states is given by the so called classical-quantum (CQ) states~\cite{Henderson2001a,Modi2012a}
	\begin{equation}
		\rho_{\rm CQ} = \sum_{i=1}^{N_A} p_i \left| i \right>_A \left<i \right| \otimes \rho_B^{(i)},
		\label{Eq:CQ-Def}
	\end{equation}
where $\{ p_{i} \}$ is a set of probabilities, ${ \{ \left| i \right>_{A} \} }$ is an orthonormal basis of $A$ and ${ \{ \rho_{B}^{(i)} \} }$ are general density matrices for subsystem $B$. Note that for any such state the existence of a commuting local observable ${ H_{A} }$ that nullifies the skew information is guaranteed (i.e. when ${ H_{A} }$ is diagonal in the basis ${ \{ \left| i \right>_{A} \} }$). The CQ states include the so called classical-classical (CC) states
	\begin{equation}
		\rho_{\rm CC} = \sum_{i=1}^{N_A} \sum_{j=1}^{N_B} p_{ij} \left| i \right>_A \left<i \right| \otimes  \left| j \right>_B \left<j \right|,
		\label{Eq:CC-Def}
	\end{equation}
where now  also ${ \{ \left| j \right>_{B} \} }$ is an orthonormal basis of $B$. Starting from an arbitrary CQ state, we plug \eq{Eq:CQ-Def} into \eq{Eq:Average-ZeroTrace} and get
	\begin{align}
		\Tr_B \left[ \left(\Tr_A \left[\sqrt{\rho_{CQ}}\right]\right)^2 \right] &= \Tr\left[ \sum_{i=1}^{N_A} p_i \rho_B^{(i)} + 2 \! \sum_{j>i=1}^{N_A} \!\!\! \sqrt{p_i p_j} \sqrt	{\rho_B^{(i)}} \sqrt{\rho_B^{(j)}} \right]  \nonumber \\
& = 1 +  2 \sum_{j>i=1}^{N_A} \sqrt{p_i p_j} \; \Tr \left[ \sqrt{\rho_B^{(i)}}\sqrt{\rho_B^{(j)}} \right].
		\label{Eq:Average-CQ-Step}																		
	\end{align}
A lower bound to \eq{Eq:Average-CQ-Step} is given by ${ \Tr_B \left[ \left(\Tr_A \left[\sqrt{\rho_{CQ}}\right]\right)^2 \right] = 1 }$. The bound is saturated, for example, when only one of the $p_i$'s is non zero, i.e. for product states $\left| \psi \right>_A \left<\psi \right| \otimes \rho_B$. Another possibility is that the $\sqrt{\rho_B^{(i)}}$'s are all orthogonal to each other. For example, the set $\{ \rho_B^{(i)} \}$ could be a set of pure orthogonal states $\{ \left| \phi_i \right>_B \left< \phi_i \right| \}$ on $B$ (thus giving a CC state).
The corresponding upper bound to the AvSk of CQ states becomes
	\begin{equation}
		\AvSk{\Lambda_{A}}{\rho_{\rm CQ}} \leq\frac{  \Tr[\Lambda_A^2 ]  }{N_A^2-1} \Big[ N_A - 1\Big].
		\label{Eq:Average-CQ-UpperBound}
	\end{equation}
We can also find an upper bound to \eq{Eq:Average-CQ-Step} if we use the inequality $\Tr\left[\left(\sqrt{\rho_B^{(i)}} - \sqrt{\rho_B^{(j)}}\right)^2\right] \geq 0$, namely
	\begin{align}
	&	1 +  2 \sum_{j>i=1}^{N_A} \sqrt{p_i p_j} \; \Tr \left[ \sqrt{\rho_B^{(i)}}\sqrt{\rho_B^{(j)}} \right] \nonumber \\ & \leq 1 +  \sum_{j>i=1}^{N_A} \sqrt{p_i p_j} \; \left( \Tr \left[ \rho_B^{(i)}\right] + \Tr \left[\rho_B^{(j)} \right] \right) \nonumber \\ & = 1 +  2 \sum_{j>i=1}^{N_A} \sqrt{p_i p_j} = \sum_{i,j=1}^{N_A} \sqrt{p_i p_j} \nonumber \\ &\leq \sum_{i,j=1}^{N_A} \frac{p_i + p_j}{2} = N_A.
		\label{Eq:Average-CQ-TraceUpperBound}
	\end{align}
The bound is saturated if and only if $p_i=1/N_A$ for each $i$ and all the $\rho_B^{(i)}$'s are equal. In other words, the CQ state must be of the form $\mathbb{I}_A/N_A \otimes \rho_B$ to have zero AvSk. In conclusion, the bounds to the AvSk of CQ states become
	\begin{equation}
		0 \leq \AvSk{\Lambda_{A}}{\rho_{\rm CQ}} \leq\frac{  \Tr[\Lambda_A^2 ]  }{N_A^2-1} \Big[ N_A - 1\Big].
		\label{Eq:Average-CQ-LowerBound}
	\end{equation}
%
%
\subsubsection*{Average skew information for quantum-classical states}
\label{Sec:Average-QC}
We can also compute the AvSk on the set of quantum-classical (QC) states that, opposite in spirit to the CQ states, can have a finite amount of discord-type quantum correlations (as measured e.g.~by the LQU with respect to subsystem $A$). An arbitrary QC state can be written as
	\begin{equation}
		\rho_{\rm QC} = \sum_{i=1}^{N_A} p_i \rho_A^{(i)} \otimes \left| i \right>_B \left<i \right| ,
		\label{Eq:QC-Def}
	\end{equation}
where ${ \{ \left| i \right>_{B} \} }$ is an orthonormal basis of $B$ and ${ \{ \rho_{A}^{(i)} \} }$ are general density matrices for subsystem $A$. We plug this into \eq{Eq:Average-ZeroTrace} and get
	\begin{align}
		1 \leq \Tr_B \left[ \left(\Tr_A \left[\sqrt{\rho_{QC}}\right]\right)^2 \right] =  \sum_{i}^{N_{B}} p_{i} \left( \Tr\left[\sqrt{\rho_A^{(i)}}\right] \right)^{2} \leq N_{A}.
		\label{Eq:Average-QC-TraceBounds}
	\end{align}
The lower bound is saturated if and only if all the $\{ \rho_A^{(i)} \}$ in \eq{Eq:QC-Def} are pure states, i.e. for all density matrices that can be written in the form 
\begin{equation}
	\rho_{\rm pQC} = \sum_{i=1}^{N_A} p_i \left| \psi_i \right>_A \left< \psi_i \right| \otimes \left| i \right>_B \left<i \right|
	\label{Eq:pQC-Def}
\end{equation}
where $ \{ \left| \psi_i \right>_A \} $ is a set of generic pure states of A (in particular we don't require them to be orthogonal, at difference with the set ${ \{ \left| i \right>_{B} \} }$). We will use the name (pure quantum)-classical (pQC) for states of the form \eq{Eq:pQC-Def}. We stress that $\rho_{\rm pQC}$ is not itself pure in general (that's why we put the word ``pure'' between parenthesis in the full name and we write a small ``p'' in the abbreviation). The upper bound is saturated if and only if all the $\{ \rho_A^{(i)} \}$ are proportional to the identity, i.e. again for states of the form $\mathbb{I}_A/N_A \otimes \rho_B$. Correspondingly, for the AvSk we get
	\begin{equation}
		0 \leq \AvSk{\Lambda_{A}}{\rho_{\rm QC}} \leq \frac{  \Tr[\Lambda_A^2 ]  }{N_A^2-1} \Big[ N_A - 1\Big].
		\label{Eq:Average-QC-bounds}
	\end{equation}
As anticipated, we see that the AvSk on the subset of QC states achieves the same bounds as the AvSk on the set of CC states. This means that general quantum correlations have no clear effect on the average susceptibility of the state. Instead, we see again that the purity of the local state of $A$ has a great importance.
%
%
\subsection{Maximum and minimum of the average skew information for general states}
\label{Sec:Average-Maximum}
Another interesting question is what states, including potentially entangled states, have the absolute highest and lowest AvSk. Consider a general bipartite state $\rho$ and its diagonal expansion onto some basis $\{ \left| \psi_{i} \right>_{AB}\}$.
	\begin{equation}
		\rho = \sum p_{i} \left| \psi_{i} \right>_{AB} \left< \psi_{i} \right|, \quad \longrightarrow \quad \sqrt{\rho} = \sum \sqrt{p_{i}} \left| \psi_{i} \right>_{AB} \left< \psi_{i} \right|.
		\label{Eq:GenericState}
	\end{equation}
From the convexity of the AvSk we have
	\begin{align}
		\AvSk{\Lambda_{A}}{\sum p_{i} \ket{\psi_{i}}_{AB} \bra{\psi_{i}}} &\leq \sum p_{i} \; \AvSk{\Lambda_{A}}{\ket{\psi_{i}}_{AB} \bra{\psi_{i}}} \nonumber \\ &\leq \frac{  \Tr[\Lambda_A^2 ]  }{N_A^2-1} \Big[ N_A - \frac{1}{\min\{N_A,N_B\}} \Big],
		\label{Eq:Average-Bounds}
	\end{align}
where the absolute maximum of the AvSk for pure states, and hence for all states, is reached only by the maximally entangled states.

We look now for the minimum. From the very definitions of the LQU and of the AvSk, we have the simple relation ${ \AvSk{\Lambda_{A}}{\rho} \geq \mathcal{U}^{\Lambda_{A}}(\rho) }$. Therefore, the states with minimum AvSk can only be found within the set of states with minimum (zero) LQU, i.e., the CQ states. As we have already seen, among all the CQ states only the states of the form $\mathbb{I}_A/N_A \otimes \rho_B$ have zero AvSk. This gives a proof of property 1b, that we formulated in Sec.~\ref{Sec:Properties}.
%
%
\subsection{Minimum of the average skew information for fixed LQU}
\label{Sec:Average-Minimum-FixedLQU}
Another interesting question is what states have AvSk equal to their LQU. We recall that the LQU can be expressed as
	\begin{equation}
		\mathcal{U}^{\Lambda_{A}}(\rho) = 1 - \Tr \left[ \sqrt{\rho} \tilde{H}_{A}  \sqrt{\rho} \tilde{H}_{A} \right],
	\end{equation}
where $\tilde{H}_{A}$ is some Hamiltonian minimizing the skew information. The AvSk over all Hamiltonians with the same spectrum can be equal to the LQU if and only if
	\begin{eqnarray*}
		\Tr \left[ \sqrt{\rho} \tilde{H}_{A}  \sqrt{\rho} \tilde{H}_{A} \right] &=& \Tr \left[ \sqrt{\rho} (U_{A}^{\dagger} \tilde{H}_{A} U_{A})  \sqrt{\rho} (U_{A}^{\dagger} \tilde{H}_{A} U_{A}) \right]  \\  &=& \Tr \left[ (U_{A} \sqrt{\rho} U_{A}^{\dagger}) \tilde{H}_{A} (U_{A}  \sqrt{\rho} U_{A}^{\dagger}) \tilde{H}_{A}  \right], \ \  \forall U_{A}. \nonumber
	\end{eqnarray*}
Note that we can always add an arbitrary local unitary transformation $V_{B}$ to the density matrix without affecting the above equality. In other words, we can ask that
	\begin{eqnarray}
		&&\Tr \left[ \sqrt{\rho} \tilde{H}_{A}  \sqrt{\rho} \tilde{H}_{A} \right]  = \Tr \left[ (U_{A} \otimes V_{B} \; \sqrt{\rho} \; U_{A}^{\dagger} \otimes V_{B}^{\dag}) \tilde{H}_{A}\right.  \nonumber \\ && \quad  \left.\times \ (U_{A} \otimes V_{B} \; \sqrt{\rho} \; U_{A}^{\dagger} \otimes V_{B}^{\dag})  \tilde{H}_{A}  \right], \quad  \forall U_{A}.
		\label{Eq:Condition}
	\end{eqnarray}
A sufficient condition for \eq{Eq:Condition} is expressed as
	\begin{equation}
		\forall U_{A}, \exists V_{B} \text{ such that } \sqrt{\rho} = (U_{A} \otimes V_{B} \; \sqrt{\rho} \; U_{A}^{\dagger} \otimes V_{B}^{\dag}).
		\label{Eq:Condition2}
	\end{equation}
Therefore, we must look for states that are invariant under any unitary operation $U_{A}$ if we allow the application of an arbitrary local correction $V_{B}$. Some examples, when ${ N_{A} = N_{B} }$, are given by the Werner states \cite{Werner1989a}, that satisfy ${ \rho_{W} = (U\otimes U) \rho_{W} (U\otimes U)^{\dagger} }$ $\forall U$, and by the isotropic states~\cite{Horodecki1999a}, that satisfy ${ \rho_{I} = (U\otimes U^{*}) \rho_{I} (U\otimes U^{*})^{\dagger}}$ $\forall U$.
%
%
\section{Average Skew Information for two qubits}
\label{Sec:Qubits}

	\begin{figure*}[thb]
		\includegraphics[trim=0pt 0pt 0pt 0pt, clip, width=0.4\textwidth]{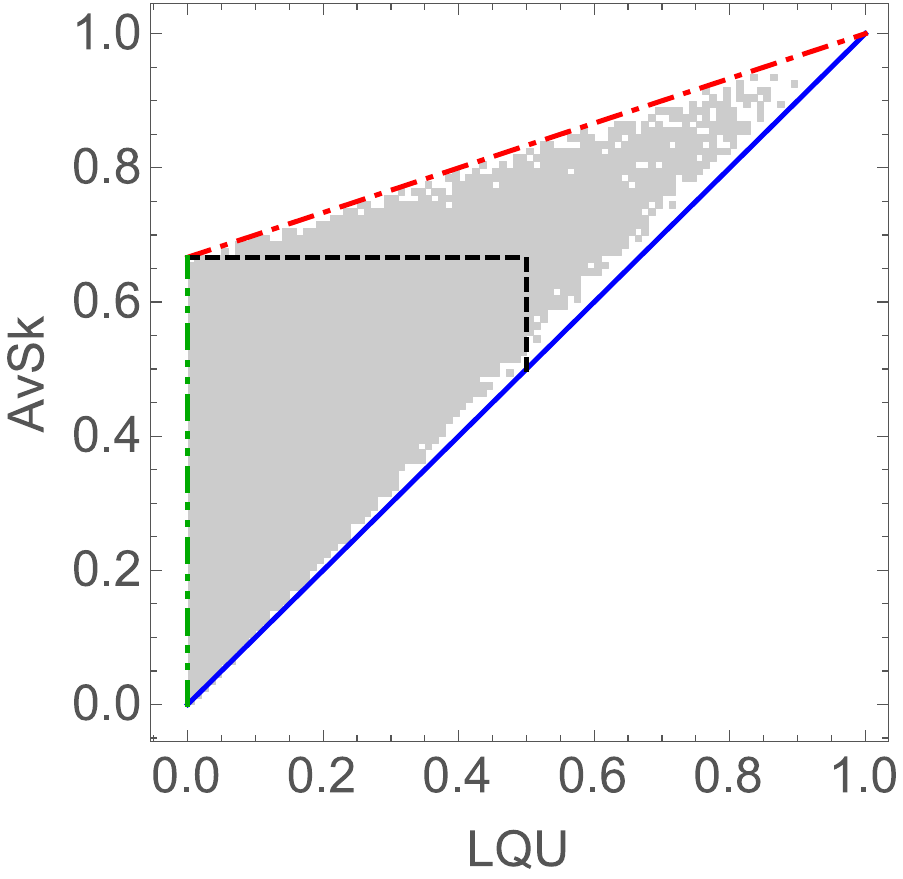}\hspace{0.05\textwidth}
		\includegraphics[trim=0pt 0pt 0pt 0pt, clip, width=0.4\textwidth]{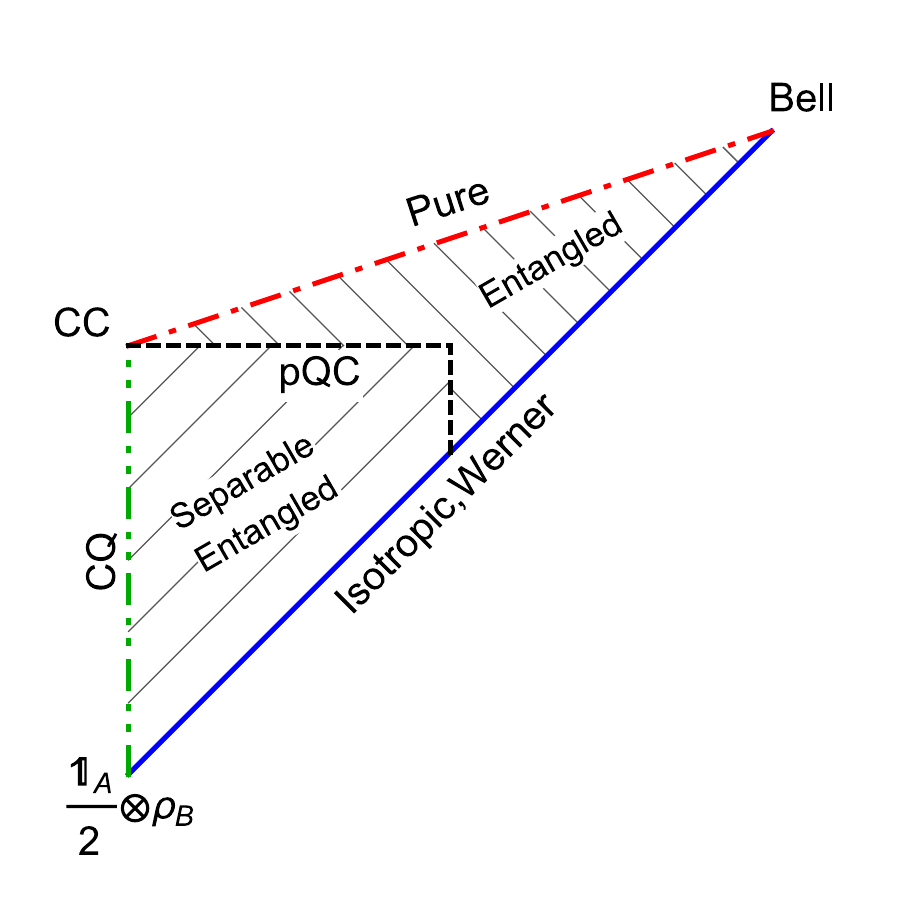}
			\caption{(Color online) AvSk and LQU for $10^5$ randomly generated two-qubit states (gray dots). Special classes of states are highlighted by different lines and detailed in the sketch on the right. See also the main text for a complete description of the various regions and boundaries.}
		\label{Fig:Numeric}
	\end{figure*}

We focus now on the exemplary case of two qubits, for which the analysis becomes particularly simple and insightful. Indeed, in this case we can also explicitly compute the LQU~\cite{Girolami2013a} and we can classify all the states according to their minimum and average susceptibility to local transformations, looking at the results of Sec.~\ref{Sec:State} in more detail. Furthermore, the LQU of two qubits coincides with their DS~\cite{Farace2014b}, and the AvSk can be then rigorously interpreted as the average discrimination efficiency of the state in a quantum illumination task~\cite{Lloyd2008a}. The analysis takes then an explicit metrological connotation.

From the results of \sec{Sec:Spectrum} we can fix ${\Lambda_{A}=\sigma_{z}}$ without loss of generality, where $\sigma_z$ is the third Pauli matrix, and the expression of the AvSk for any two-qubit state becomes then
	\begin{equation}
		\AvSk{\sigma_{z}}{\rho} =\frac{2}{3} \Bigg[ 2 - \Tr_B\Big[(\Tr_{A}[\sqrt{\rho}])^2\Big]\Bigg].
		\label{Eq:Average-ZeroTrace-Qubits}
	\end{equation}
We compute the AvSk and the LQU (using the formula in~\cite{Girolami2013a}) for $10^5$ randomly generated two-qubit states. In Fig.~\ref{Fig:Numeric} we plot the AvSk of each state vs. the corresponding LQU.

The results of~\sec{Sec:State} are clearly illustrated by the plot. Namely, we observe the following:

\begin{itemize}
\item
 Since the LQU is obtained through a minimization over all possible unitaries and the AvSk is obtained through an average, we must have that ${ \AvSk{\sigma_{z}}{\rho} \geq \mathcal{U}^{\sigma_{z}}(\rho) }$. This lower bound, shown by a blue solid line in Fig.~\ref{Fig:Numeric}, is saturated, for example, by isotropic and Werner states (see Appendix~\ref{App:Isotropic}).

\item The separable states, including the CQ states (for which the LQU vanishes) and the QC states, satisfy the bound $\left\{  \AvSk{\sigma_{z}}{\rho_{\rm sep}} ,  \AvSk{\sigma_{z}}{\rho_{\rm CQ}} ,  \AvSk{\sigma_{z}}{\rho_{\rm QC}} \right\} \leq 2/3 $. CQ states are shown by a green dot-dot-dashed line in Fig.~\ref{Fig:Numeric}. $\rm{pQC}$ states have all ${   \AvSk{\sigma_{z}}{\rho_{\rm pQC}} = 2/3 }$ and are shown by the horizontal dashed black line. CC states all have $\mathcal{U}^{\sigma_{z}}(\rho_{\text{CC}})=0$ and ${   \AvSk{\sigma_{z}}{\rho_{\rm CC}} = 2/3 }$.

\item From Ref.~\cite{Farace2014b}, we know that the separable states must have limited LQU,  $\mathcal{U}^{\sigma_{z}}(\rho_{\text{sep}}) \leq 1/2$. Therefore, they must lie left of the vertical dashed black line in Fig.~\ref{Fig:Numeric}. Combined with the previous observation, this allows us to identify a region where only entangled states exist and a region where separable and entangled states coexist (see also appendix~\ref{App:Isotropic}).

\item The pure states satisfy the bound ${ 2/3 \leq  \AvSk{\sigma_{z}}{\ket{\psi}_{AB}} \leq 1 }$. The lower bound is saturated by separable pure states and the upper bound is saturated by maximally entangled states (Bell states). The Bell states also achieve the highest AvSk among all states.
Moreover, for any given value of the LQU, the highest possible AvSk is achieved by a pure state. For pure states of two qubits, we have
	\begin{equation}
		\AvSk{\sigma_{z}}{\ket{\psi}_{AB}} =\frac{2}{3}\Bigg[ 1 + \frac{\mathcal{U}^{\sigma_{z}}(\ket{\psi}_{AB})}{2} \Bigg] = \frac{2}{3} \Bigg[ 1 + \frac{C(\ket{\psi}_{AB})^{2}}{2} \Bigg],
		\label{Eq:Average-Conc2}
	\end{equation}
where $C(\ket{\psi}_{AB})$ is the concurrence. Pure states are indicated by a red dot-dashed line in Fig.~\ref{Fig:Numeric}.

\item States of the form ${ (\mathbb{I}_{A}/2) \otimes \rho_{B} }$ are the only states having zero AvSk.
\end{itemize}

The simultaneous analysis of the AvSk and of the LQU provides a useful guide when we need to decide which states of the two-qubit probe are more suitable to perform a given metrological task (e.g., in the present case, state discrimination). We immediately see that maximally entangled states, as can be expected, are the best choice when we focus on both the worst-case performance and the average performance as figures of merit. However, if we have limited resources and do not have access to entangled states, we can still achieve good results using separable states. For example, the state 
\begin{equation}\label{Eq:MaxDiscordant}
\tilde{\rho} = \frac{1}{2} \ket{0}_{A} \bra{0} \otimes \ket{0}_{B} \bra{0} +  \frac{1}{2} \ket{+}_{A} \bra{+} \otimes \ket{1}_{B} \bra{1}
\end{equation} yields a LQU equal to $1/2$ and an AvSk equal to $2/3$ (recall that the maximum is $1$ for both quantities). This state, among all separable states, has the highest amount of discord-type correlations. This confirms that quantum correlations beyond entanglement are indeed useful for metrological applications, although they play a relevant role only in determining the worst-case performance but have little effect on the average performance (the value $2/3$ for the AvSk can be reached even with product states). Another observation that we can make is the following. If one needs to guarantee a minimum efficiency of the probe, i.e. fix the LQU as a primary figure of merit, there is still some freedom in the choice of the initial state, with pure states being on average better than any other possibility. Our analysis of the AvSk can be very useful in this sense.

%
%
\section{Variance of the skew information}
\label{Sec:Variance}
In this Section, we complement the above analysis by computing the variance of the skew information, which tells us how much the efficiency of a given probe state is fluctuating around the average value for different choices of the encoding unitary. The variance is defined as
\begin{eqnarray}
		\Delta \AvSk{\Lambda_{A}}{\rho} = \int d \mu_H (U_A) I^{2}(\rho, U_A \Lambda_A {U_A}^\dagger)  - \left( \AvSk{\Lambda_{A}}{\rho} \right)^{2}.
	\label{Eq:Variance}
	\end{eqnarray}
The first term on the right-hand side of \eq{Eq:Variance} can be computed following the prescriptions of Appendices~\ref{App:Zuber} and~\ref{App:Variance}, and the expression for two-qubit states is given by Eq.~\eqref{Eq:SecondMomentSolve}. We can further impose that $\Lambda_{A} = \sigma_{z}$ without loss of generality. Since analytic insight is out of reach for such a cumbersome expression, we resort again to computing the variance numerically for the $10^{5}$ random two-qubit states generated before.

	\begin{figure*}[htbp]
		{\includegraphics[trim=0pt 0pt 100pt 0pt, clip, height=0.35\textwidth]{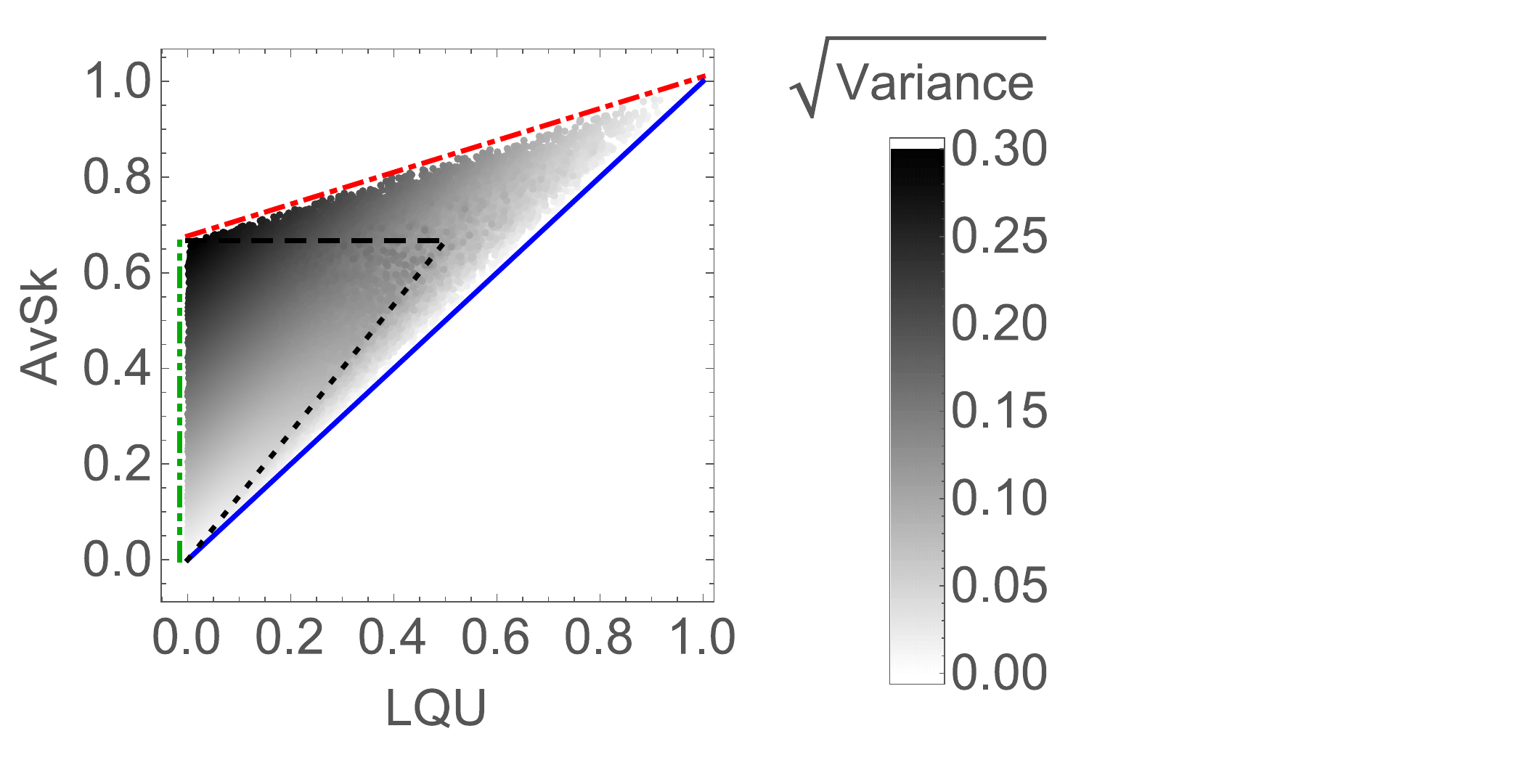}}
		{\includegraphics[trim=0pt -15pt 0pt -15pt, clip, height=0.35\textwidth]{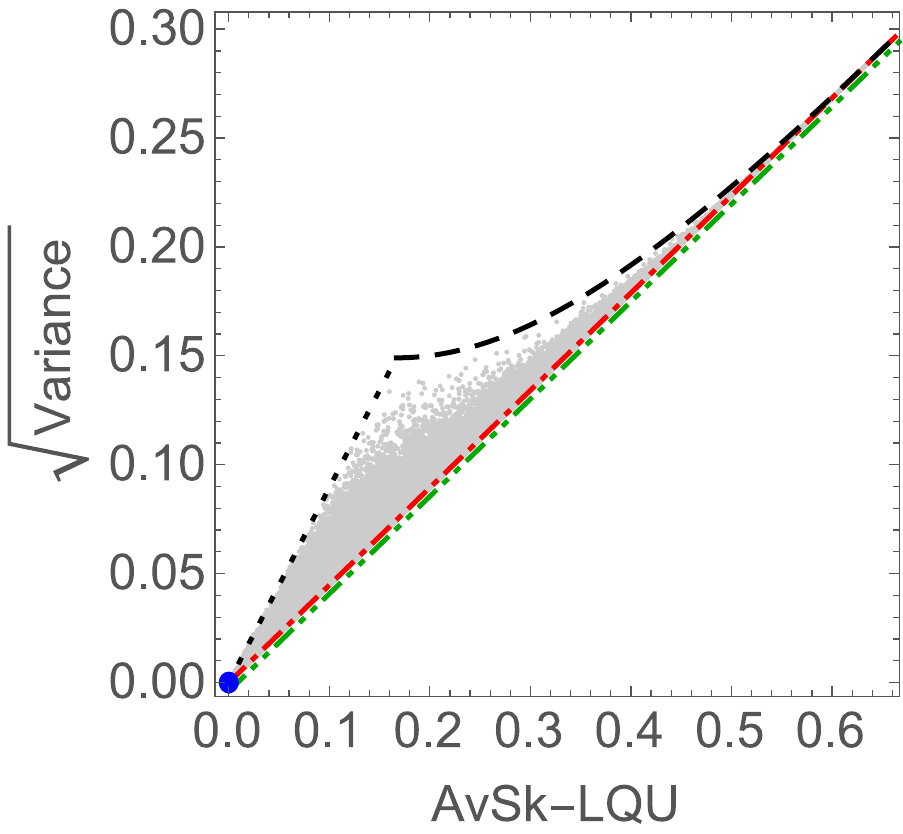}}
			\caption{(Color online) Left: Color plot of the square root of the variance of the skew information for $10^5$ randomly generated two-qubit states (gray dots), as a function of the AvSk and of the LQU. Right: Square root of the variance of the skew information plotted as a function of ${AvSk - LQU}$. Special classes of states (detailed in the main text) are highlighted by different lines, using the same style in both figures for comparison.}
		\label{Fig:NumericVariance}
	\end{figure*}

The results are presented in Fig.~\ref{Fig:NumericVariance}. On the left panel, we show a density plot of the square root of the variance, given the corresponding AvSk and LQU. It is immediate to see that the variance is zero on the isotropic and Werner states, for which the average is equal to the minimum, and is bigger when the difference between the average (AvSk) and the minimum (LQU) is bigger, as should be expected. Moreover, we find that there are precise quantitative relations that describe this behavior. To show this, on the right panel of Fig.~\ref{Fig:NumericVariance} we plot the square root of the variance versus the corresponding value of the difference (AvSk $-$ LQU). All the points lie within a well-defined region and we can find the states lying on the boundaries by constructing educated guesses based on several special classes of states that are simple to parameterize.
%
%
\subsection{Pure states}
\label{Sec:QubitsVariancePure}
We expand the pure states of two qubits in their Schmidt basis and write them as ${\kets{\psi}{AB} = \sum_{i=1}^{2} \sqrt{c_{i}} \left| i \right>_{A} \left| i \right>_{B}}$, where ${ \{ \left| i \right>_{A}\} }$ and ${ \{ \left| i \right>_{B} \} }$ are orthonormal basis of $A$ and $B$ and ${ c_{1} + c_{2}=1 }$, so they are easily parameterized by a single number $c_{1}$. Thanks to this, we can easily show that
	\begin{equation}
		\AvSk{\sigma_{z}}{\ket{\psi}_{AB}}-\mathcal{U}^{\sigma_{z}}(\ket{\psi}_{AB})=\frac{2}{3}(1-2 c_1)^{2}.
	\end{equation}
Moreover, Eq.~\eqref{Eq:SecondMomentSolve} can be greatly simplified for pure states and the variance can be computed analytically,
	\begin{equation}
		\Delta \AvSk{\sigma_{z}}{\ket{\psi}_{AB}} = \frac{4}{45}(1-2 c_1)^{4}.
		\label{Eq:VarianceSkewPure}
	\end{equation}	
In the end, we find the simple relation
	\begin{equation}
		\sqrt{\Delta \AvSk{\sigma_{z}}{\ket{\psi}_{AB}} }= \frac{1}{\sqrt{5}} \Big( \AvSk{\sigma_{z}}{\ket{\psi}_{AB}}-\mathcal{U}^{\sigma_{z}}(\ket{\psi}_{AB}) \Big).
		\label{Eq:VarianceSkewPureRelation}
	\end{equation}	
These states provide the lower boundary for the right plot of Fig.~\ref{Fig:NumericVariance} and are highlighted with a red dot-dashed line. A red dot-dashed line is also shown in the left plot for comparison.
%
%
\subsection{Product states}
\label{Sec:QubitsVarianceProduct}
Next we consider the product states, which all have zero LQU. We seek a family of product states depending on only one parameter and interpolating between one state of the form ${ (\mathbb{I}_{A}/2) \otimes \rho_{B} }$, which has zero AvSk, and one pure product state, which has the highest AvSk among product states. Therefore, we consider the family of states given by
	\begin{equation}
		\rho_{\rm prod}(p) := \Big( p \left| 0 \right>_{A}\left< 0 \right| + (1-p) \frac{\mathbb{I}_{A}}{2} \Big) \otimes \left| 0 \right>_{B}\left< 0 \right|, \text{ with } p \in [0,1] .
		\label{Eq:FamilyProduct}
	\end{equation}
Their AvSk is easily computed	
	\begin{equation}
		\AvSk{\sigma_{z}}{\rho_{\rm prod}(p)}=\frac{2}{3}\left( 1 -\sqrt{1-p^{2}} \right).
	\end{equation}
Eq.~\eqref{Eq:SecondMomentSolve} can be again evaluated analytically in this case and becomes
	\begin{equation}
		 \Delta \AvSk{\sigma_{z}}{\rho_{\rm prod}(p)} = \frac{4}{45}\left( 1 -\sqrt{1-p^{2}} \right)^{2}.
		\label{Eq:VarianceSkewPure}
	\end{equation}	
In the end, we find the same simple relation as for pure states,
	\begin{equation}
		\sqrt{\Delta \AvSk{\sigma_{z}}{\rho_{\rm prod}(p)} } = \frac{1}{\sqrt{5}} \Big( \AvSk{\sigma_{z}}{\rho_{\rm prod}(p)}-\mathcal{U}^{\sigma_{z}}(\rho_{\rm prod}(p)) \Big).
		\label{Eq:VarianceSkewProductRelation}
	\end{equation}	
Therefore, these product states also lie on the lower boundary of the right plot of Fig.~\ref{Fig:NumericVariance} and are highlighted with a green dot-dot-dashed line. A green dot-dot-dashed line is also shown in the left plot for comparison.
%
%
\subsection{Pure-QC states}
\label{Sec:QubitsVariancePureQC}
Another interesting class of states is given by the (pure quantum)-classical states, introduced above (see \eq{Eq:pQC-Def} in Sec.~\ref{Sec:Average-Product}). For two qubits, these are states of the form
	\begin{equation}
		\rho_{\rm pQC} = p \ketbras{\psi_{0}}{\psi_{0}}{A} \otimes \ketbras{0}{0}{B} + (1-p) \ketbras{\psi_{1}}{\psi_{1}}{A} \otimes \ketbras{1}{1}{B},
		\label{Eq:PureQC}
	\end{equation}	 
with $\left| \psi_{0} \right>_{A}$ and $\left| \psi_{1} \right>_{A}$ arbitrary pure states. A special subset of pure-QC states is obtained by taking
	\begin{equation}
		\rho_{\rm pQC}(p) = \frac{1-p}{2} \ketbras{0}{0}{A} \otimes \ketbras{0}{0}{B} + \frac{1+p}{2} \ketbras{+}{+}{A} \otimes \ketbras{1}{1}{B},
		\label{Eq:PureQC}
	\end{equation}	
with $p \in [0,1]$, which linearly interpolates between a pure product state (when $p=1$) and the maximally discordant separable state of Eq.~(\ref{Eq:MaxDiscordant}), i.e. the one having LQU equal to $1/2$~\cite{Farace2014b} (when $p=0$). These states have constant AvSk, with value $2/3$. Their LQU and their variance can also be explicitly computed as functions of $p$, although we do not report here the expressions. Combining the three quantities, we get the following relation
	\begin{eqnarray}
&&		\sqrt{ \Delta \AvSk{\sigma_{z}}{\rho_{\rm pQC}(p)} } \\ && = \frac{1}{3\sqrt{5}} \sqrt{ 1 + 3 \left( \Big( \AvSk{\sigma_{z}}{\rho_{\rm pQC}(p)}-\mathcal{U}^{\sigma_{z}}(\rho_{\rm pQC}(p))  \Big) - \frac{1}{3} \right)^{2} }. \nonumber
		\label{Eq:VarianceSkewPureQCRelation}
	\end{eqnarray}	
We conjecture that pure-QC states provide the upper boundary for the right plot of Fig.~\ref{Fig:NumericVariance} as highlighted with a black dashed line. This  is well supported by the numerical evidence. A black dashed line is also shown in the left plot for comparison.
%
%
\subsection{Separable states}
\label{Sec:QubitsVarianceSeparable}
From \sec{Sec:QubitsVariancePureQC} we see that the state which behaves most differently (in terms of the variance) with respect to pure and product states is given by the maximally discordant separable state of Eq.~(\ref{Eq:MaxDiscordant}). The leftmost upper curve for the right plot of Fig.~\ref{Fig:NumericVariance} (shown by a black dotted line) connects this state $\tilde{\rho}$ with a state of the form ${ (\mathbb{I}_{A}/2) \otimes \rho_{B} }$. We then make an ansatz that separable states of the form
	\begin{equation}
		\rho_{\rm sep}(p) = \frac{p}{2} \left(\ketbras{0}{0}{A} \otimes \ketbras{0}{0}{B} +  \ketbras{+}{+}{A} \otimes \ketbras{1}{1}{B}\right) + \left(1 - p \right) \frac{\mathbb{I}}{4},
		\label{Eq:SeparableP}
	\end{equation}	
with $p \in [0,1]$, will attain the boundary. We see that once again our ansatz is well supported by the numerics. A black dotted line is also shown in the left plot for comparison.  For states of the form given in \eq{Eq:SeparableP}, the relation between the variance, the average and the LQU is given by
	\begin{equation}
		\sqrt{\Delta \AvSk{\sigma_{z}}{\rho_{\rm sep}(p)} } = \frac{2}{\sqrt{5}} \Big( \AvSk{\sigma_{z}}{\rho_{\rm sep}(p)}-\mathcal{U}^{\sigma_{z}}(\rho_{\rm sep}(p)) \Big).
		\label{Eq:VarianceSkewSeparable}
	\end{equation}	
%
%
\section{The role of correlations}
\label{Sec:Correlations}
In this last Section we are going to discuss the influence played by correlations on the average and the variance of the skew information.
%
%
\subsection{Bounds on quantum correlations}
\label{Sec:Bounds}
We have already seen that the amount of quantum (discord-type) correlations has no specific effect on the AvSk. However, since the minimum susceptibility (i.e. the LQU) is instead a proper measure of quantum correlations, the variance of the skew information is affected as well. Roughly speaking, we can see that if two states have the same AvSk but one state has more discord-type correlations (i.e. higher LQU), its variance will be smaller compared to the other. We can turn this into a quantitative statement and derive bounds for the LQU by combining the information about the average and the variance of the skew information. In the case of two qubits, these bounds read
	\begin{eqnarray}
&&		\max \left\{0, \AvSk{\sigma_{z}}{\rho}-\sqrt{5 \Delta \AvSk{\sigma_{z}}{\rho}} \right\} \leq \mathcal{U}^{\sigma_{z}}(\rho)  \\  && \quad \leq \begin{cases}
		\AvSk{\sigma_{z}}{\rho} - \frac{1}{2} \sqrt{ 5 \Delta \AvSk{\sigma_{z}}{\rho} } \quad & \text{if } \Delta \AvSk{\sigma_{z}}{\rho} \leq 1/45;\\
		\AvSk{\sigma_{z}}{\rho} - \frac{1}{3} - \sqrt{15 \Delta \AvSk{\sigma_{z}}{\rho} - \frac{1}{3}} \quad & \text{if } 1/45 < \Delta \AvSk{\sigma_{z}}{\rho}. \nonumber
		\end{cases}
		\label{Eq:Bounds}
	\end{eqnarray}
We expect that these bounds become tighter and tighter by introducing higher moments of the statistics. Indeed, knowing more about the distribution should also give more information about the minimum. For two-qubit states this is not very useful, since we can easily compute the LQU directly \cite{Girolami2013a}. However, in higher dimensions it is not known how to perform analytically the optimization involved in the computation of the LQU and one has to resort to numerical analysis. The approach presented here exploits quantities (the average and the variance of the skew information) that are exactly computable in any dimension (although their expressions could be rather involved) and could be then easily generalized beyond the two-qubit case.
%
%
\subsection{Connections with a recent measure introduced by Luo et al.}
\label{Sec:Luo}
We now show that the AvSk can be corrected to yield a proper measure of correlations. In order to do this we recall that a big contribution to the value of the AvSk comes from the purity of the local state of subsystem $A$. Roughly speaking, this accounts for the fact that a pure state $\ket{\psi_{A}}$ is more efficient than a mixed state (e.g. the maximally mixed state $\rho_{A} = \mathbb{I}_{A}/N_{A}$) in detecting the action of local operations. Moreover, if we think in terms of the discrimination protocol described in~\cite{Farace2014b}, we can consider the situation in which we use a bipartite probe but perform only local measurements on $A$ to extract the information (e.g. if we lose access to the reference subsystem $B$). In the latter case, the only resource that we can exploit is the local purity of the state $\rho_{A} = \Tr_{B} [\rho_{AB}]$. Every benefit that we gain by measuring the global state $\rho_{AB}$ must hence come from correlations. Motivated by the above reasoning, we define the quantity
 	\begin{eqnarray} 		\label{Eq:AvSkCorr}
		\AvSkCorr{\Lambda_{A}}{ \rho_{AB}} &\equiv &\AvSk{\Lambda_{A}}{ \rho_{AB}}  - \AvSk{\Lambda_{A}}{\rho_A \otimes \rho_B}  \\  &=& \frac{ \Tr[\Lambda_A^2 ] }{N_A^2-1} \Bigg[  \left(\Tr_{A} \left [\sqrt{\Tr_{B} [\rho_{AB}] } \right] \right)^2 - \Tr_B\Big[(\Tr_{A}[\sqrt{\rho_{AB}}])^2\Big]\Bigg] . \nonumber
	\end{eqnarray}	
Note that the quantity  $\AvSk{\Lambda_{A}}{ \rho_A \otimes \rho_B}$ in \eq{Eq:AvSkCorr} actually depends only on $\rho_A$ and not on the other marginal $\rho_B=\Tr_{A} [\rho_{AB}]$.

We now show that the above quantity is equal (up to a prefactor) to the quantity introduced by Luo and collaborators in~\cite{Luo2012a}. They define
	\begin{equation}
		F(\rho_{AB}) \equiv Q_{A} (\rho_{AB}) - Q_{A} (\rho_A \otimes \rho_B),
		\label{Eq:LuoCorr}
	\end{equation}	
where $Q_{A} (\rho)$ is the average of the skew information with respect to any family of $N_{A}^{2}$ orthonormal hermitian operators, i.e. with respect to any orthonormal basis for the real Hilbert space $L(\mathcal{H}_{A})$ according to the scalar product $\langle A, B \rangle = \Tr [A B]$. That is,
	\begin{equation}
		Q_{A} (\rho) = \sum_{i=1}^{N_{A}^{2}} I(\rho,X_{i}), \quad \text{with } X_{i} \in L(\mathcal{H}_{A}) \text{ and } \Tr[X_{i} X_{j}] = \delta_{ij}.
		\label{Eq:LuoDef}
	\end{equation}	
The quantity $Q_{A}(\rho)$ can be evaluated by writing the skew information as in \eq{Eq:Skew-swap} and noting that $\sum_{i=1}^{N_{A}^{2}} X_{i} \otimes X_{i} = S_{A|A'}/N_{A}$, where $S_{A|A'}$ is the swap operator. After some manipulations, the final expression $Q_{A} (\rho_{AB}) = N_{A} - \Tr_B\Big[(\Tr_{A}[\sqrt{\rho_{AB}}])^2\Big]$ can be found. We see that $Q_{A}(\rho_{AB})$ coincides with $\AvSk{\Lambda_{A}}{ \rho_{AB}}$ apart from a numerical prefactor that depends only on the choice of the spectrum. Therefore $\AvSkCorr{\Lambda_{A}}{ \rho_{AB}}$ is proportional to $F(\rho_{AB})$ in general. Luo et al. have shown \cite{Luo2012a} that $F(\rho_{AB})$  satisfies the following properties:

- $F(\rho_{AB}) = 0$ if and only if $\rho_{AB} = \rho_{A} \otimes \rho_{B}$.

- $F(\rho_{AB}) = F(U_{A} \otimes V_{A} \rho_{AB} U_{A}^{\dag} \otimes V_{A}^{\dag})$ is invariant under local unitary operations, $\forall$ $U_{A},$ $V_{B}$.

- $F(\rho_{AB})$ is decreasing under  arbitrary CPTP maps on subsystem $B$. It is also conjectured that $F(\rho_{AB})$ is decreasing under arbitrary CPTP maps on subsystem $A$.

From these, they argue that $F(\rho_{AB})$ is a measure of {\it total} correlations, but cannot be specifically regarded as a measure of classical or quantum correlations. We conclude that the same holds for our quantity $\AvSkCorr{\Lambda_{A}}{ \rho_{AB}}$ defined in Eq.~(\ref{Eq:AvSkCorr}).

Our analysis, though, complements the results of Luo et al. in two key points. First, we have provided a closed and compact expression that can be evaluated for any dimension of the Hilbert spaces $\mathcal{H}_{A}$ and $\mathcal{H}_{B}$ and that is not explicitly found in~\cite{Luo2012a}. Second, we have given a clear operative interpretation to $\AvSkCorr{\Lambda_{A}}{ \rho_{AB}}$, as the advantage that we can gain in doing some metrology task (e.g. state discrimination) by fully exploiting the amount of correlations in a bipartite probe.
%
%
\section{Skew information and metrology}
\label{Sec:Metrology}
As we discussed in the introduction and mentioned several times throughout the paper, the choice of the skew information has the twofold advantage of allowing easy manipulations and retaining interesting connections with the field of quantum metrology. Here we make these connections explicit, in the hope of conveying a clearer message to the reader before moving to the conclusions. 

Two common problems that are studied in quantum metrology are phase estimation~\cite{Helstrom1976a,Huelga1997a,Giovannetti2006a,Escher2011a,Giovannetti2011a} and state discrimination~\cite{Helstrom1976a,Audenaert2007a,Calsamiglia2008a}. In phase estimation, the goal is estimating a continuos parameter $\phi$ that characterises the unitary transformation ${ \rho \rightarrow e^{-i H \phi} \rho e^{i H \phi} }$ of the initial state of the probe. After choosing the best possible measurement strategy and the best possible estimator $\tilde \phi_{best}$ (i.e. a function of the measurement outcomes and probabilities that is used to guess the actual value $\phi$), the achievable precision in the limit of $n \gg 1$ repetitions of the protocol is determined by the quantum Cramer-Rao bound~\cite{Helstrom1976a,Braunstein1994a}, which relates the variance of the estimator $\Delta^2 \tilde \phi_{best} \stackrel{n \gg 1}{=} 1/(n F(\rho,H))$ to the inverse of the quantum Fisher information $F(\rho,H)$. In state discrimination, the goal is discriminating between the initial state of the probe $\rho$ and a transformed state $\rho'$. Since $\rho$ and $\rho'$ are in general not orthogonal, therefore not perfectly distinguishable, the figure of merit used in this case is the probability $p$ of guessing correctly, that scales asymptotically in the number $n$ of repetitions as $p(n) \stackrel{n \gg 1}{\sim} 1-Q(\rho,\rho')^n$, where $Q(\rho,\rho') = \min_s \Tr[\rho^s \rho'^{1-s}]$ is the quantum Chernoff bound~\cite{Audenaert2007a,Calsamiglia2008a}.

The skew information is strictly related to both the quantum Fisher information and to the quantum Chernoff bound, as we show in the following. First of all, the original quantum Fisher information~\cite{Helstrom1976a,Braunstein1994a} is not the only generalization of the classical Fisher information, but there is a whole family of so-called generalized quantum Fisher informations~\cite{Gibilisco2003a,Toth2014a}. They all share a set of fundamental properties, e.g. are convex and have the same value for pure states, and they are all upper bounded by the original quantum Fisher information. The Wigner-Yanase skew information (multiplied by a factor $4$) belongs to this family~\cite{Toth2014a} and this fact combined with another recent result~\cite{Luo2004a}, allows us to write  
	\begin{equation}
		\frac{1}{ 8 I[\rho,H]} \leq n \Delta^2 \tilde \phi_{best} \stackrel{n \gg 1}{=} \frac{1}{F[\rho,H]} \leq \frac{1}{ 4 I[\rho,H]}.
	\end{equation}
We see that the skew information can be used to set upper and lower bounds to the estimation precision. Therefore, if we fix a set of isospectral generators $H(\Lambda)$ for the unitary phase transformation $e^{-i H(\Lambda) \phi}$, the LQU~\cite{Girolami2013a} and the AvSk give strong indications about the minimum and the average estimation precision with respect to this set.

Second, it was shown in~\cite{Farace2014b} that the quantity $1 - Q(\rho,\rho')$ shares strong connections with the skew information. Specifically $I(\rho,H)$ can be seen as the efficiency of a discrimination process where the two states that need to be distinguished are given by ${ \{\rho, e^{-iH} \rho e^{iH} \} }$  and the unitary transformation $e^{-iH}$ is a small perturbation of the identity operator. That is, we have the relation 
	\begin{equation}
		I(\rho,H) \stackrel{| h_i | \ll 1, \forall i}{\sim} 1 - Q(\rho, e^{-iH} \rho e^{iH})
	\end{equation}
where $\{ h_i \}$ are the eigenvalues of $H$ (see~\cite{Farace2014b} for a formal characterisation). Moreover, the relation between the skew information and the quantum Chernoff bound becomes even more stringent when $H$ is any operator acting on the Hilbert space of a qubit: indeed, in this special case the two quantities are proportional and we get
	\begin{equation}
		I(\rho,H) \stackrel{H = a \mathbb I_2 + \vec b \cdot \vec \sigma }{\propto} 1 - Q(\rho, e^{-iH} \rho e^{iH}).
	\end{equation}	
We see that the LQU and the AvSk can therefore be used to characterise the minimum and average efficiency in discriminating the elements of any of the couples $\{ \rho, e^{-iH_1} \rho e^{iH_1} \}$, $\dots$, $\{ \rho, e^{-iH_n} \rho e^{iH_n} \}$, where $H_1, \dots, H_n$ belong to a set of isospectral Hamiltonians. 

All the above discussion remains valid even if we assume that $\rho = \rho_{AB}$ is a bipartite state and the transformations act only on subsystem $A$, as we did throughout the paper. Moreover, with these additional assumptions we can use the LQU and the AvSk to draw another bridge between quantum metrology and quantum information theory, analysing the role of several resources in enhancing the metrological performance of different quantum states. It's precisely in this sense that the analysis of \sec{Sec:Qubits} acquires a strong relevance with respect to metrological applications.
%
%
\section{Conclusions}
\label{Sec:Conclusions}

In this work we have investigated a question of joint fundamental and practical relevance, namely which resources and which bipartite states are useful as versatile probes, to achieve a required average performance in quantum metrology tasks \cite{Giovannetti2006a,Giovannetti2011a} involving a variable set of operations encoding an unknown parameter on one subsystem. We demonstrated that the average susceptibility of a state to local unitary transformations can be reliably quantified by the average skew information, a quantity that we introduce and calculate in closed form for bipartite quantum states of arbitrary dimension. The average skew information is found to be a convex measure strongly dependent on the local purity of the probing subsystem, and in general requiring entanglement to reach its maximum value. However, separable or even product states can still achieve fairly satisfactory degrees of average skew information, meaning that they can be reliable metrological resources on average, when entanglement is not available.
  
The results of our analysis have been contrasted with the related, but different setting, in which the worst-case (rather than the average) susceptibility to local transformations is studied for bipartite states \cite{Girolami2013a,Girolami2014a,Adesso2014a,Farace2014b,Roga2015a}. Such a worst-case performance can be quantified by the minimum skew information, known as local quantum uncertainty \cite{Girolami2013a}, which is instead determined entirely by quantum correlations of the discord type.
By analysing comparatively the minimum, the average, and the variance of the skew information, we have identified the role of state purity, separability, and correlations to identify probe states with extremal properties, classifying their broad potential for metrological tasks such as parameter estimation and state discrimination. The general analysis has been illustrated in particular in the simplest yet particularly relevant instance of two-qubit probe states, for which we have provided a complete numerical characterization.

In this paper we were not concerned with another important issue in quantum metrology, i.e. how the precision of the estimation scales as we increase the number of ``constituents'' in the probe. Much is already known on the problem. For example, as mentioned in the introduction, one can show that by using pure entangled states of $n$ qubits the minimum estimation error can be reduced by a factor $\sqrt{n}$ with respect to using a pure separable state~\cite{Giovannetti2006a,Giovannetti2011a}. Moreover, a recent work~\cite{Modi2011a} has provided evidence that a similar enhancement can be found for discordant mixed states of $n$ qubits over classical mixed states, under particular measurement strategies. We remark that in this paper we didn't study the role of correlations (and other properties of the probe) with respect to optimal performances. Instead we focused on a complementary aspect, i.e. versatility, and left outside, at least  for the moment, considerations regarding the scaling of our functionals. 

We expect that the study of the minimum and average skew information for continuous variable systems~\cite{Ferraro2005a,Adesso2007a,Weedbrook2012a} would provide us with further insights In this case, for example, it comes naturally that one does not have experimental access to the whole infinite-dimensional Hilbert space and must work with limited resources (e.g. limited classes of states and operations, limited energy, limited squeezing, limited amount of entanglement, limited purity, ...). The minimum and the average skew information would provide then clear and simple-to-evaluate criteria that can help in picking optimal probes among the set of accessible states. Based on the recent progress in calculating some of these measures (such as the interferometric power and the discriminating strength) for Gaussian states of continuous variable systems in a worst-case scenario \cite{Adesso2014a,Roga2015a,Rigovacca2015a}, we believe that a Gaussian version of the average skew information might be amenable to analytical evaluation; it would then become particularly important to study its scaling with the resources typically involved in optical interferometry, such as the mean energy of the probing system~\cite{Adesso2014a}, and with other nonclassical features such as squeezing and entanglement. This is left for future investigation.

This work has provided yet another application of the Wigner-Yanase skew information, defined more than half a century ago \cite{Wigner1963a}, in quantum information theory. The skew information represents one of the most insightful and mathematically convenient quantum generalizations of the classical Fisher information \cite{Gibilisco2003a,Toth2014a}, and it has proven useful already to derive improved uncertainty relations \cite{Luo2003a}, to define measures of asymmetry (coherence) \cite{Girolami2014b,Marvian2014a} and correlations \cite{Luo2012a,Girolami2013a}, and to construct generalized geometric quantum speed limits \cite{Paiva2015a}. The latter application, in particular, deals with the question: How fast can a quantum state evolve under a closed or open system evolution? The study presented in this work can be framed in a similar perspective, as the average skew information introduced here quantifies precisely how fast, on average, a quantum state of a bipartite system evolves under any local unitary dynamics (within a fixed spectral class) affecting one of its subsystems. The more versatile probes for quantum metrology are exactly those whose reaction to the local dynamics is faster, indicating an increased susceptibility to the unknown parameter encoded in the dynamics itself.

It is finally interesting to comment on the information-theoretic resource unlocking such an enhanced susceptibility to local dynamics. If the figure of merit is the minimum susceptibility, the resource is local asymmetry (coherence) in all possible reference bases for the probing subsystem, which is equivalent to discord-type quantum correlations \cite{Girolami2013a,Girolami2014a}. If the figure of merit is the average susceptibility, instead, we demonstrated that the resource is local purity for the probing subsystem. This is clear when one considers that the ``free'' states with vanishing average skew information are those of Eq.~(\ref{Eq:FreeStates}), taking the form of a product of the maximally mixed state for the probing subsystem, tensor any state for the other reference subsystem. Therefore any degree of local purity becomes useful in this context. This suggests that the average skew information could be further investigated as a quantum thermodynamical resource \cite{Horodecki2013a}. Namely, considering the case in which the probing system $A$ has all degenerate energy levels (so that the maximally mixed local states are the only free states), the average skew information defined in this paper might be related to the amount of work that can be extracted from  $A$ by some optimal thermal machine with access to the reference storage system $B$, provided the machine is coupled to a heat bath \cite{Faist2015a}. Investigating these intriguing connections further will be the subject of an independent study.

%
%
\section*{Acknowledgments}
\label{Sec:Acknowledgments}
We thank Thomas Bromley, Giacomo De Palma, Jonathan Oppenheim, Renato Renner, and Andreas Winter for fruitful discussions. This work was supported by the EU Collaborative Project TherMiQ (Grant agreement 618074) and the European Research Council (ERC StG GQCOP, Grant No.~637352).
%
%
\newpage
\begin{widetext}
\begin{appendix}
%
%
\section{The swap operator}
\label{App:Swap}
Let $\mathcal{H}_X$ and $\mathcal{H}_{X'}$ be isomorphic Hilbert spaces spanned by the orthonormal basis $\{\kets{i}{X}\}$ and $\{\kets{i}{X'}\}$. The $swap$ or $flip$ operator $S_{X|X'}$ is defined by the relation $S_{X|X'}\kets{i}{X} \kets{j}{X'}=\kets{j}{X} \kets{i}{X'}$ \cite{Brun2001a}. A possible representation is given by
\begin{equation}
S_{X|X'} = \sum_{i,j} \ketbras{i}{j}{X} \otimes  \ketbras{j}{i}{X'}\;.
\end{equation}
We report some useful properties that we use throughout the paper.
\begin{enumerate}
\item $(S_{X|X'})^{2}=\mathbb{I}_{XX'}$, where $\mathbb{I}_{XX'}=\mathbb{I}_{X}\otimes\mathbb{I}_{X'}$ is the identity
 operator on $\mathcal{H}_X \otimes\mathcal{H}_{X'}$;
 \item $S_{X|X'}=S_{A|A'} \otimes S_{B|B'}$, if $\mathcal{H}_X=\mathcal{H}_A \otimes \mathcal{H}_B$ and $\mathcal{H}_{X'}=\mathcal{H}_{A'} \otimes \mathcal{H}_{B'}$,  with $\mathcal{H}_{A}$ isomorphic to $\mathcal{H}_{A'}$ and $\mathcal{H}_{B}$ isomorphic to $\mathcal{H}_{B'}$;
 \item $S_{X|X'} (\Theta_X \otimes \Omega_{X'})S_{X|X'} =(\Omega_X \otimes \Theta_{X'})$, where $\Theta_{X,X'}$ and $\Omega_{X,X'}$ are linear operators on the corresponding Hilbert spaces;
 \item $(\Theta_X\otimes \Omega_{X'})S_{X|X'}=S_{X|X'}(\Omega_X\otimes \Theta_{X'})$, which simply follows from Property 3 if we apply $S_{X|X'}$ to both terms;
 \item $\Tr_{X,X'}[(\Theta_X\otimes \Omega_{X'})S_{X|X'}]=\Tr_{X}[\Theta_X \Omega_X]$.\\
\end{enumerate}
For completeness, we sketch the proof of Property 5. Without loss of generality we set
\begin{equation}
\Theta_X=\sum_{i,j} \theta_{ij}\ketbras{i}{j}{X}, \quad \Omega_X=\sum_{\ell,m} \omega_{\ell m}\ketbras{\ell}{m}{X}\,.
\end{equation}
By explicit computations we have
\begin{equation}
\Tr_{X}[\Theta_X \Omega_X] = \sum_{ij} \theta_{ij} \omega_{ji} \;,
\end{equation}
and
\begin{eqnarray}
\Tr_{X,X'}[(\Theta_X\otimes \Omega_{X'})S_{X|X'}]&=&\sum_{ij}\sum_{\ell,m} \sum_{\alpha,\beta}\theta_{ij} \omega_{\ell m} \Tr[\ketbras{i}{j}{X}\ketbrasbis{\alpha}{\beta}{X} \otimes \ketbras{\ell}{m}{X'}\ketbrasbis{\beta}{\alpha}{X'}] \nonumber\\
&=&\sum_{ij}\sum_{\ell,m} \sum_{\alpha,\beta}\theta_{ij} \omega_{\ell m} \, \delta_{\alpha j}\delta_{\beta i}\delta_{\alpha \ell} \delta_{\beta m}=\sum_{ij}\sum_{\ell,m} \theta_{ij} \omega_{\ell m} \, \delta_{j \ell} \delta_{i m}=\sum_{ij} \theta_{ij} \omega_{ji} \;,
\end{eqnarray}
thus concluding the proof.
%
%
\section{The twirling channel}
\label{App:Twirling}
Let $\mathcal{H}_X$ and $\mathcal{H}_{X'}$ be isomorphic Hilbert spaces and $\Theta_{XX'}$ an operator acting on the tensor of the two ${\mathcal{H}_X \otimes \mathcal{H}_{X'}}$. The twirling channel modifies this operator by applying the same local unitary operation simultaneously to $X$ and $X'$ and then averaging this action over all possible local unitaries. This is expressed as
	\begin{eqnarray}
		{\cal T}^{(2)} (\Theta_{XX'} ) =  \int d \mu_H(U) \; \left( U_{X} \otimes U_{X'}\right)\; \Theta_{XX'} \;
( U_{X}^\dagger \otimes U_{X'}^\dagger)\,,
		\label{Eq:Twirling-def}
	\end{eqnarray}
where $d \mu_H (U)$ is the Haar measure over the unitary group $\{ U(N_{X}) \}$ and ${N_X= \dim ({\cal H}_X)}$. It can be shown that the integral in \eq{Eq:Twirling-def} has a simple solution in terms of the swap operator as introduced in Appendix~\ref{App:Swap}
	\begin{eqnarray}
  		{\cal T}^{(2)} (\Theta_{XX'} ) \ug \tfrac{N_X \Tr [\Theta_{XX'}]  - \Tr [ S_{X|X'} \Theta_{XX'}]  }{N_X (N_X^2-1)} \; \mathbb{I}_{XX'}
+ \tfrac{N_X  \Tr [S_{X|X'} \Theta_{XX'}]-\Tr [\Theta_{XX'}]}{N_X (N_X^2-1)}\;  S_{X|X'}  \;.
		\label{Eq:Twirling-res}
	\end{eqnarray}

%
%
\section{Isotropic states of two qubits}
\label{App:Isotropic}
We consider the isotropic states of two qubits~\cite{Horodecki1999a}
	\begin{equation}
		\rho_F= \frac{1-F}{3} \mathbb{I}_{AB} + \frac{4F-1}{3} \ketbras{\psi_+}{\psi_+}{},
	\end{equation}
parametrized by $0 \leq F \leq 1$, where $F$ is the fidelity between the isotropic state and the Bell state $\left| \psi_{+} \right> = (\left| 00 \right> + \left| 11 \right>)/\sqrt{2}$. The AvSk can be easily computed by decomposing the identity on the Bell basis.
	\begin{equation}
		\rho_F =  \frac{1-F}{3} \ketbras{\phi_-}{\phi_-}{} + \frac{1-F}{3} \ketbras{\phi_+}{\phi_+}{} + \frac{1-F}{3} \ketbras{\psi_-}{\psi_-}{} + F \ketbras{\psi_+}{\psi_+}{},
	\end{equation}
so that
	\begin{equation}
		\sqrt{\rho_F} =  \sqrt{\frac{1-F}{3}} \ketbras{\phi_-}{\phi_-}{} + \sqrt{\frac{1-F}{3}} \ketbras{\phi_+}{\phi_+}{} + \sqrt{\frac{1-F}{3}} \ketbras{\psi_-}{\psi_-}{} + \sqrt{F} \ketbras{\psi_+}{\psi_+}{}.
	\end{equation}
Plugging this into Eq. \eqref{Eq:Average-ZeroTrace}, we get that the AvSk is given by ${ 1- \left[ 2 \tfrac{1-F}{3} + 2 \sqrt{F}\sqrt{\tfrac{1-F}{3}} \right] }$. Similarly, the LQU can be computed following the prescription of~\cite{Girolami2013a} and the result is again given by the same expression ${ 1- \left[ 2 \tfrac{1-F}{3} + 2 \sqrt{F}\sqrt{\tfrac{1-F}{3}} \right] }$, as anticipated in the main text. Isotropic states of two qubits lie on the blue line in Fig.~\ref{Fig:Numeric}. Entangled isotropic states (with $F\geq 1/2$) have AvSk=LQU$\geq \tfrac{2-\sqrt{3}}{3} \sim 0.09$. Therefore, this gives a simple proof that some entangled states lie on the left of the curve LQU$=1/2$.
%
%
\section{Integrals over the unitary group}
\label{App:Zuber}
Consider a general integral of the form
	\begin{eqnarray}
		\int d \mu_H(U) \; U_{i_{1},j_{1}} U_{i_{2},j_{2}} \dots U_{i_{n},j_{n}} \; (U^{\dagger})_{k_{1},\ell_{1}} (U^{\dagger})_{k_{2},\ell_{2}} \dots (U^{\dagger})_{k_{m},\ell_{m}}\,,
		\label{Eq:UnitaryMoments}
	\end{eqnarray}
where $U$ is a unitary matrix acting on the Hilbert space ${ {\cal H}_X }$, ${ d \mu_H (U) }$ is the Haar measure over the unitary group $ \{ U(N_{X}) \} $ and ${N_X= \dim ({\cal H}_X)}$. Such integral is called a moment of order $(n,m)$ of the unitary group and is zero whenever ${ n \neq m }$~\cite{Collins2006a}. In the case ${ n = m }$, the integral can be computed using the Weingarten calculus~\cite{Weingarten1978a} and yields the expression~\cite{Zuber2008a}
	\begin{eqnarray}
		\sum_{\sigma, \tau \in S_{n}} c (n, \sigma) \prod_{a=1}^{n} \delta_{i_{a} \ell_{\tau(a)}} \delta_{j_{a} k_{\tau \sigma(a)}} \; ,
		\label{Eq:UnitaryMomentsFormal}
	\end{eqnarray}
where $\sigma$ and $\tau$ belong to the symmetric group, i.e. they are permutations of $n$ elements, and $c(n, \sigma)$ are the so called Weingarten functions which depend on the number of elements appearing in the integral and on the particular permutation of those n elements. The analytic expression of the Weingarten functions is explicitely known for small values of $n$~\cite{Zuber2008a}, and it can be computed for higher $n$ with some effort.

Note that the twirling channel implicitly contains an integral of the form~\eqref{Eq:UnitaryMoments}, with ${ n=m=2 }$.

\subsection{Weingarten functions for the case $n=m=4$}
\label{App:Case}
The case $n=m=4$ is particularly interesting to us since it appears in the computation of the variance of the skew information. We report here the Weingarten functions for $n=4$, taking them from~\cite{Zuber2008a}. First let us set some notation to deal with permutations. A permutation $\sigma$ of $4$ elements will be written as its action on the string ${ \{1,2,3,4\} }$. So for example, the permutation ${ ( 3 \; 2 \; 4 \; 1 ) }$ maps ${ \{1,2,3,4\} }$ to ${ \{3,2,4,1\} }$, i.e. brings the first element to the fourth place, the third element to the first place, the fourth element to the third place and leaves the second element unchanged. Where possible we will index these permutation by the associated permutation class, e.g. ${ ( 3 \; 2 \; 4 \; 1 ) } \to [\sigma]=[1,3]$ as given by one cycle over 3 elements ($1$, $3$ and $4$ in the example) and one cycle over one element ($2$ in the example).  With this in mind, we can now write
	\begin{equation}
		\begin{array}{c|c}
		\sigma & c(4,\sigma,N_{X})\\ \hline \\
		\begin{minipage}{0.22\textwidth} \vspace{5pt} $[\sigma]=[4]$
\vspace{5pt} \end{minipage}    & \dfrac{-5}{(N_{X}-3)(N_{X}-2)(N_{X}-1)N_{X}(N_{X}+1)(N_{X}+2)(N_{X}+3)}\\ \\
		  \hline \\
		\begin{minipage}{0.22\textwidth} \vspace{5pt} $[\sigma]=[1,3]$
		 \vspace{5pt} \end{minipage} & \dfrac{2N_{X}^{2}-3}{(N_{X}-3)(N_{X}-2)(N_{X}-1)N_{X}(N_{X}+1)(N_{X}+2)(N_{X}+3)}\\ \\ \hline \\
		\begin{minipage}{0.22\textwidth} \vspace{5pt} $[\sigma]=[2^2]$
		\vspace{5pt} \end{minipage}& \dfrac{N_{X}^{2}+6}{(N_{X}-3)(N_{X}-2)(N_{X}-1)N_{X}(N_{X}+1)(N_{X}+2)(N_{X}+3)}\\ \\ \hline \\
		\begin{minipage}{0.22\textwidth} \vspace{5pt} $[\sigma]=[1^2,2]$
		 \vspace{5pt} \end{minipage} & \dfrac{1}{(N_{X}-3)(N_{X}-1)N_{X}(N_{X}+1)(N_{X}+3)}\\ \\ \hline \\
		\begin{minipage}{0.22\textwidth} \vspace{15pt} $[\sigma]=[1^4]$
		 \vspace{15pt} \end{minipage} & \dfrac{N_{X}^{4}-8N_{X}^{2}+6}{(N_{X}-3)(N_{X}-2)(N_{X}-1)N_{X}(N_{X}+1)(N_{X}+2)(N_{X}+3)}
		\end{array}
		\label{Eq:Weingarten4}
	\end{equation}	
In other words, the Weingarten functions depend only on the class of the permutation. An important thing to notice is that the Weingarten functions with $n=4$ diverge for ${ N_{X}<4 }$ (and in general they diverge when ${ N_{X}<n }$). However, it has been proven that the sum in Eq.~\eqref{Eq:UnitaryMomentsFormal} does not diverge because the poles in each term cancel out after careful simplifications~\cite{Collins2006a}. This allows us to compute the variance of the skew information even for two-qubit states (having ${ N_{X}=2 }$).
%
%
\section{Second moment of the skew information for two-qubit states}
\label{App:Variance}
The second moment of the skew information is defined as
	\begin{equation}
		\left< I^{2}(\rho, \Lambda_A) \right>_{\{U_{A}\}}:=\frac{1}{4} \int d \mu_H (U_A) \left( \Tr\left[ [\sqrt{\rho},U_A \Lambda_A {U_A}^\dagger]^2\right] \right)^{2}.
		\label{Eq:SecondMoment}
	\end{equation}
This expression can be expanded as
	\begin{align}		
		\int d \mu_H (U_A) & \Big( \Tr[\sqrt{\rho} U_{A} \Lambda^{2}_A U_{A}^\dagger \sqrt{\rho}]^{2} + \Tr[\sqrt{\rho} U_{A} \Lambda_A U_{A}^\dagger \sqrt{\rho} U_{A} \Lambda_A U_{A}^\dagger]^{2} \nonumber \\
		& \qquad - 2 \; \Tr[\sqrt{\rho} U_{A} \Lambda_A U_{A}^\dagger U_{A} \Lambda_A U_{A}^\dagger \sqrt{\rho}] \; \Tr[\sqrt{\rho} U_{A} \Lambda_A U_{A}^\dagger \sqrt{\rho} U_{A} \Lambda_A U_{A}^\dagger] \Big).
	\label{Eq:SecondMomentExpand}
	\end{align}
	
The first term in Eq.~\eqref{Eq:SecondMomentExpand} can be integrated using the properties of the twirling channel and the result reads
	\begin{align}		
		\int d \mu_H (U_A) \; \Tr[\sqrt{\rho} U_{A} \Lambda^{2}_A U_{A}^\dagger \sqrt{\rho}]^{2} = \frac{2 \Tr[ \Lambda_{A}^{2} ]^{2} - \Tr[ \Lambda_{A}^{4} ]}{6} + \frac{2 \Tr[ \Lambda_{A}^{4} ] - \Tr[ \Lambda_{A}^{2} ]^{2}}{6} \Tr_{A}[ (\Tr_{B}[\rho])^{2}].
	\label{Eq:SecondMoment1}
	\end{align}
	
The second term in Eq.~\eqref{Eq:SecondMomentExpand} can be rewritten by expanding each operator on a basis ${ \{ \left| i \right>_{A} \left| j \right>_{B} \} }$ of $\cal{H}_{AB}$. For simplicity we fix the local basis of $A$ to make $\Lambda_{A}$ diagonal.
	\begin{align}		
		&\int d \mu_H (U_A) \; \Tr[\sqrt{\rho} U_{A} \Lambda_A U_{A}^\dagger \sqrt{\rho} U_{A} \Lambda_A U_{A}^\dagger]^{2} = \nonumber \\
		&\int d \mu_H (U_A) \Big( \sum_{ij} \sqrt{\rho}_{i_{1} j_{1}, i_{2} j_{2}} U_{i_{2},i_{3}} \Lambda_{i_{3}} U^{\dagger}_{i_{3},i_{4}} \sqrt{\rho}_{i_{4} j_{2}, i_{5}, j_{1}} U_{i_{5}, i_{6}} \Lambda_{i_{6}} U^{\dagger}_{i_{6},i_{1}} \nonumber \\
		& \qquad \qquad \times \sum_{i'j'} \sqrt{\rho}_{i'_{1} j'_{1}, i'_{2} j'_{2}} U_{i'_{2},i'_{3}} \Lambda_{i'_{3}} U^{\dagger}_{i'_{3},i'_{4}} \sqrt{\rho}_{i'_{4} j'_{2}, i'_{5}, j'_{1}} U_{i'_{5}, i'_{6}} \Lambda_{i'_{6}} U^{\dagger}_{i'_{6},i'_{1}} \Big) = \nonumber \\
		&  \sum_{iji'j'} \sqrt{\rho}_{i_{1} j_{1}, i_{2} j_{2}} \sqrt{\rho}_{i_{4} j_{2}, i_{5}, j_{1}} \sqrt{\rho}_{i'_{1} j'_{1}, i'_{2} j'_{2}} \sqrt{\rho}_{i'_{4} j'_{2}, i'_{5}, j'_{1}} \Lambda_{i_{3}}  \Lambda_{i_{6}} \Lambda_{i'_{3}} \Lambda_{i'_{6}} \nonumber \\
		 &  \qquad \qquad \times \int d \mu_H (U_A) \Big( U_{i_{2},i_{3}}  U_{i_{5}, i_{6}}    U_{i'_{2},i'_{3}}  U_{i'_{5}, i'_{6}}  U^{\dagger}_{i_{3},i_{4}} U^{\dagger}_{i_{6},i_{1}}  U^{\dagger}_{i'_{3},i'_{4}}  U^{\dagger}_{i'_{6},i'_{1}} \Big).
		\label{Eq:SecondMoment2}
	\end{align}
Using the results of Appendix~\ref{App:Zuber}, we can compute the integral over the unitary matrices and \eq{Eq:SecondMoment2} reduces to
	\begin{align}		
		\sum_{\sigma \tau} c(4,\sigma, 2) \; F( \tau \sigma, \Lambda) \; G_{2}(\tau, \rho)
		\label{Eq:SecondMoment2Solve}
	\end{align}
where
		\begin{align}		
		F(\tau \sigma, \Lambda_{A}) = \sum_{i_{3} i_{6} i'_{3} i'_{6}}\Lambda_{i_{3}}  \Lambda_{i_{6}} \Lambda_{i'_{3}} \Lambda_{i'_{6}} \delta_{\{ i_{3}, i_{6}, i'_{3}, i'_{6} \}, \tau \sigma (\{ i_{3}, i_{6}, i'_{3}, i'_{6} \}) }
		\label{Eq:SecondMomentF}
	\end{align}
is a function which depends only on the spectrum and on the composition of permutations $\tau \sigma$, while
	\begin{align}		
		G_{2}(\tau, \rho) = \sum_{i_{1}i_{2}i_{4}i_{5}}  \sum_{i'_{1}i'_{2}i'_{4}i'_{5}}  \sum_{j_{1}j_{2}}  \sum_{j'_{1}j'_{2}} \sqrt{\rho}_{i_{1} j_{1}, i_{2} j_{2}} \sqrt{\rho}_{i_{4} j_{2}, i_{5}, j_{1}} \sqrt{\rho}_{i'_{1} j'_{1}, i'_{2} j'_{2}} \sqrt{\rho}_{i'_{4} j'_{2}, i'_{5}, j'_{1}} \delta_{\{ i_{2}, i_{5}, i'_{2}, i'_{5} \}, \tau ( \{ i_{4}, i_{1}, i'_{4}, i'_{1} \} )}
		\label{Eq:SecondMomentG2}
	\end{align}
is a function which depends only on the state and on the permutation $\tau$. Both $F$ and $G_{2}$ can be analytically computed for each choice of $\sigma$ and $\tau$. Assuming that the spectrum is traceless, we find that $F$ depends only on the class on the permutation $\tau \sigma$
	\begin{equation}
		\begin{array}{c|c}
		\tau \sigma & F(\tau \sigma, \Lambda_{A})\\ \hline
		\begin{minipage}{0.44\textwidth} \vspace{5pt} $[\tau\sigma]=[4]$
		\vspace{5pt} \end{minipage}  & \Tr[ \Lambda_{A}^{4} ] \\ \hline
		\begin{minipage}{0.44\textwidth} \vspace{5pt} $[\tau\sigma]=[1,3]$
		\vspace{5pt} \end{minipage} & 0 \\ \hline
		\begin{minipage}{0.44\textwidth} \vspace{5pt} $[\tau\sigma]=[2^2]$
		\vspace{5pt} \end{minipage}& \Tr[ \Lambda_{A}^{2} ]^{2} \\ \hline
		\begin{minipage}{0.44\textwidth} \vspace{5pt} $[\tau\sigma]=[1^2,2]$
		\vspace{5pt} \end{minipage} & 0\\ \hline
		\begin{minipage}{0.44\textwidth} \vspace{5pt} $[\tau\sigma]=[1^4]$
		\vspace{5pt} \end{minipage} & 0
		\end{array}
		\label{Eq:TableF}
	\end{equation}	
while $G_{2}$ assumes different values even among permutations belonging to the same class
	\begin{equation}
		\begin{array}{c|c}
		\tau & G_{2}(\tau, \rho) \\ \hline
		( 1 \; 2 \; 3 \; 4 ) & 1  \\ \hline
		( 1 \; 2 \; 4 \; 3 ) & \mathcal{A}  \\ \hline
		( 1 \; 3 \; 2 \; 4 ) & \mathcal{B}  \\ \hline
		( 1 \; 3 \; 4 \; 2 ) & \mathcal{E}  \\ \hline
		( 1 \; 4 \; 2 \; 3 ) & \mathcal{E}  \\ \hline
		( 1 \; 4 \; 3 \; 2 ) & \mathcal{B}  \\ \hline
		( 2 \; 1 \; 3 \; 4 ) & \mathcal{A}  \\ \hline
		( 2 \; 1 \; 4 \; 3 ) & \mathcal{C}  \\ \hline
		\end{array} \hspace{20pt}
		\begin{array}{c|c}
		\tau & G_{2}(\tau, \rho) \\ \hline
		( 2 \; 3 \; 1 \; 4 ) & \mathcal{E} \\ \hline
		( 2 \; 3 \; 4 \; 1 ) & \mathcal{F}  \\ \hline
		( 2 \; 4 \; 1 \; 3 ) & \mathcal{F}  \\ \hline
		( 2 \; 4 \; 3 \; 1 ) & \mathcal{E}  \\ \hline
		( 3 \; 1 \; 2 \; 4 ) & \mathcal{E}  \\ \hline
		( 3 \; 1 \; 4 \; 2 ) & \mathcal{F} \\ \hline
		( 3 \; 2 \; 1 \; 4 ) & \mathcal{B}  \\ \hline
		( 3 \; 2 \; 4 \; 1 ) & \mathcal{E}  \\ \hline
		\end{array} \hspace{20pt}
		\begin{array}{c|c}
		\tau & G_{2}(\tau, \rho) \\ \hline
		( 3 \; 4 \; 1 \; 2 ) & \mathcal{D}  \\ \hline
		( 3 \; 4 \; 2 \; 1 ) & \mathcal{G} \\ \hline
		( 4 \; 1 \; 2 \; 3 ) & \mathcal{F}  \\ \hline
		( 4 \; 1 \; 3 \; 2 ) & \mathcal{E}  \\ \hline
		( 4 \; 2 \; 1 \; 3 ) & \mathcal{E}  \\ \hline
		( 4 \; 2 \; 3 \; 1 ) & \mathcal{B}  \\ \hline
		( 4 \; 3 \; 1 \; 2 ) & \mathcal{G}  \\ \hline
		( 4 \; 3 \; 2 \; 1 ) & \mathcal{D} \\ \hline
		\end{array}
		\label{Eq:TableG2}
	\end{equation}	
The various letters in the last table are shorthand notation for the following expressions
	\begin{align}
		&\mathcal{A} = \Tr_{B} \left[ (\Tr_{A}[ \sqrt{\rho} ])^{2} \right], \nonumber \\
		&\mathcal{B} = \Tr_{A} \left[ (\Tr_{B}[ \rho])^{2} \right], \nonumber \\
		&\mathcal{C} = \mathcal{A}^{2}, \nonumber \\
		&\mathcal{D} = \Tr_{AA'} \left[ ( \Tr_{B} [ (\mathbb{I}_{A'} \otimes \sqrt{\rho_{AB}}) \cdot (\mathbb{I}_{A} \otimes \sqrt{\rho_{A'B}}) ] )^{2} \right], \nonumber \\
		&\mathcal{E} = \Tr \left[ \sqrt{\rho} \cdot (\Tr_{B}[ \rho ] \otimes \Tr_{A}[ \sqrt{\rho} ] ) \right], \nonumber \\
		&\mathcal{F} = \Tr_{A} \left[ (\Tr_{B}[ \sqrt{\rho} \cdot \left( \mathbb{I}_{A} \otimes \Tr_{A}[ \sqrt{\rho} ] \right) ])^{2} \right], \nonumber \\
		&\mathcal{G} = \Tr_{AA'} \left[ ( \Tr_{B} [ (\mathbb{I}_{A'} \otimes \sqrt{\rho_{AB}}) \cdot (\mathbb{I}_{A} \otimes \sqrt{\rho_{A'B}}) ] )^{2} S_{A|A'}\right].
	\label{Eq:Letters}
	\end{align}
In the last expression, $S_{A|A'}$ is the swap operator \cite{Brun2001a} discussed in Appendix~\ref{App:Swap}.

The third term in Eq.~\eqref{Eq:SecondMomentExpand} can be tackled similarly to the second term (note that we explicitly wrote it with $U_{A}$ and $U_{A}^{\dagger}$ appearing 4 times each). The result is
	\begin{align}		
		-2 \sum_{\sigma \tau} c(4,\sigma, 2) \; F( \tau \sigma, \Lambda) \; G_{3}(\tau, \rho)
		\label{Eq:SecondMoment3Solve}
	\end{align}
where $F$ is the defined in Eq.~\eqref{Eq:TableF} and $G_{3}$ is defined below
	\begin{equation}
		\begin{array}{c|c}
		\tau & G_{3}(\tau, \rho) \\ \hline
		( 1 \; 2 \; 3 \; 4 ) & 1  \\ \hline
		( 1 \; 2 \; 4 \; 3 ) & \mathcal{A}  \\ \hline
		( 1 \; 3 \; 2 \; 4 ) & \mathcal{B}  \\ \hline
		( 1 \; 3 \; 4 \; 2 ) & \mathcal{E}  \\ \hline
		( 1 \; 4 \; 2 \; 3 ) & \mathcal{E}  \\ \hline
		( 1 \; 4 \; 3 \; 2 ) & \mathcal{B}  \\ \hline
		( 2 \; 1 \; 3 \; 4 ) & 2  \\ \hline
		( 2 \; 1 \; 4 \; 3 ) & 2 \mathcal{A}  \\ \hline
		\end{array} \hspace{20pt}
		\begin{array}{c|c}
		\tau & G_{3}(\tau, \rho) \\ \hline
		( 2 \; 3 \; 1 \; 4 ) & 1 \\ \hline
		( 2 \; 3 \; 4 \; 1 ) & \mathcal{A}  \\ \hline
		( 2 \; 4 \; 1 \; 3 ) & \mathcal{A}  \\ \hline
		( 2 \; 4 \; 3 \; 1 ) & \mathcal{A}  \\ \hline
		( 3 \; 1 \; 2 \; 4 ) & 2 \mathcal{B}  \\ \hline
		( 3 \; 1 \; 4 \; 2 ) & 2 \mathcal{E} \\ \hline
		( 3 \; 2 \; 1 \; 4 ) & \mathcal{B}  \\ \hline
		( 3 \; 2 \; 4 \; 1 ) & \mathcal{E}  \\ \hline
		\end{array} \hspace{20pt}
		\begin{array}{c|c}
		\tau & G_{3}(\tau, \rho) \\ \hline
		( 3 \; 4 \; 1 \; 2 ) & \mathcal{E}  \\ \hline
		( 3 \; 4 \; 2 \; 1 ) & \mathcal{B} \\ \hline
		( 4 \; 1 \; 2 \; 3 ) & 2 \mathcal{E}  \\ \hline
		( 4 \; 1 \; 3 \; 2 ) & 2 \mathcal{B}  \\ \hline
		( 4 \; 2 \; 1 \; 3 ) & \mathcal{E}  \\ \hline
		( 4 \; 2 \; 3 \; 1 ) & \mathcal{B}  \\ \hline
		( 4 \; 3 \; 1 \; 2 ) & \mathcal{B}  \\ \hline
		( 4 \; 3 \; 2 \; 1 ) & \mathcal{E} \\ \hline
		\end{array}
		\label{Eq:TableG3}
	\end{equation}	
By putting together Eqs.~\eqref{Eq:SecondMoment1},~\eqref{Eq:SecondMoment2Solve} and~\eqref{Eq:SecondMoment3Solve} we find
	\begin{align}
		\left< I^{2}(\rho, \Lambda_A) \right>_{\{U_{A}\}} & =\frac{2 \Tr[ \Lambda_{A}^{2} ]^{2} - \Tr[ \Lambda_{A}^{4} ]}{6} + \frac{2 \Tr[ \Lambda_{A}^{4} ] - \Tr[ \Lambda_{A}^{2} ]^{2}}{6} \mathcal{B} \nonumber \\
		& + \sum_{\sigma \tau} c(4,\sigma, 2) \; F( \tau \sigma, \Lambda) \; \big(G_{2}(\tau, \rho) -2G_{3}(\tau, \rho) \big).
		\label{Eq:SecondMomentSolve}
	\end{align}
Unfortunately, we cannot further simplify this expression to explicitly show that the poles appearing in each $c(4,\sigma, 2)$ cancel out. However, a direct computation proves that Eq.~\eqref{Eq:SecondMomentSolve} does not diverge.
%
%

\end{appendix}
%
%
\clearpage
\end{widetext}
\newpage
\bibliography{average-bib}
\bibliographystyle{apsrev4-1}
%
%

%

\end{document}